\documentclass[11pt]{article}

\usepackage[margin=1in]{geometry}
\usepackage{mathtools}
\usepackage{amssymb}
\usepackage{amsthm}
\usepackage{bm}
\usepackage{amsmath,booktabs}

\usepackage{booktabs,array,microtype}
\usepackage[colorlinks=true,linkcolor=blue,urlcolor=blue]{hyperref}
\usepackage[x11names]{xcolor}

\hypersetup{
  citecolor=SpringGreen4
}

\usepackage{enumitem}
\usepackage{needspace}

\usepackage{mathtools}
\mathtoolsset{showonlyrefs=true,showmanualtags=true}

\usepackage[bottom]{footmisc}
\usepackage{tablefootnote}\usepackage{diagbox}
\usepackage{cellspace} 
\newcommand{\AppendixBanner}{%
  \hbox to \textwidth{%
    \leaders\hbox{\rule[.55ex]{1em}{0.5pt}}\hfill
    \enspace\textbf{APPENDIX}\enspace
    \leaders\hbox{\rule[.55ex]{1em}{0.5pt}}\hfill
  }%
}

\makeatletter
\def\ps@appendix{%
  \let\@mkboth\markboth
  \def\@oddhead{\AppendixBanner}%
  \def\@evenhead{\AppendixBanner}%
  \def\@oddfoot{\hfil\thepage\hfil}%
  \def\@evenfoot{\hfil\thepage\hfil}%
}
\makeatother

\usepackage{tikz}
\usetikzlibrary{calc}
\usetikzlibrary{decorations.pathreplacing}

\usepackage{pgfplots}
\pgfplotsset{compat=1.17}

\newtheorem{theorem}{Theorem}[section]
\newtheorem{lemma}[theorem]{Lemma}

\newtheorem{corollary}[theorem]{Corollary}

\theoremstyle{definition}

\theoremstyle{remark}

\newcommand{\bu}{\bm{u}}
\newcommand{\bv}{\bm{v}}

\newcommand{\cC}{{\mathcal C}}
\newcommand{\eps}{\varepsilon}

\newcommand{\Exp}{\mathbb{E}}
\renewcommand{\Pr}{\mathbb{P}}

\newcommand{\EQ}{\;=\;}
\newcommand{\LE}{\;\le\;}
\newcommand{\LT}{\;<\;}
\newcommand{\GE}{\;\ge\;}

\newcommand{\Eq}{\;\!\!=\;\!}

\newcommand{\ds}{\displaystyle}

\newcommand{\OPT}{\mathrm{OPT}}
\newcommand{\SDP}{\mathrm{SDP}}

\title{No polynomial-time $\frac78$-approximation algorithm\\ for MAX SAT, assuming UGC and $\mathsf{P} \neq \mathsf{NP}$}

\author{Joshua Brakensiek\thanks{University of California, Berkeley. Email: \texttt{josh.brakensiek@berkeley.edu}. Supported in part by Venkatesan Guruswami's Simons Investigator award and NSF grants CCF-2211972 and DMS-2503280.}\and Neng Huang\thanks{TU Dresden. NH has been funded by the European Union (Project POCOCOP, ERC Synergy Grant 101071674). Views and opinions expressed are however those of the author(s) only and do not necessarily reflect those of the European Union or the European Research Council Executive Agency. Neither the European Union nor the granting authority can be held responsible for them. Email: \texttt{neng.huang@tu-dresden.de}}\and Aaron Potechin\thanks{University of Chicago. Supported in part by NSF grant CCF-2611927. Email: \texttt{potechin@uchicago.edu}}\and Uri Zwick\thanks{Blavatnik School of Computer Science and AI, Tel Aviv University, Israel. Email: \texttt{zwick@tau.ac.il}. Work supported by ISF grant 2735/2025.}}

\begin{document}

\newcommand{\ratiofour}{0.8749}
\newcommand{\ratiofive}{0.8742}
\newcommand{\rationine}{0.8734}

\maketitle

\null\vspace*{-0.75cm}
\begin{abstract}
    We show that the integrality ratio of the standard SDP relaxation of the MAX $k$-SAT problem, where clauses can have up to $k$ variables, is strictly below $\frac78$ for every $k\ge 5$. Using a result of Raghavendra, and assuming the Unique Games Conjecture (UGC) and that $\mathsf{P} \neq \mathsf{NP}$, this implies that there is no polynomial-time $\frac78$-approximation algorithm for MAX $k$-SAT, for every $k\ge 5$, and hence for the MAX SAT problem in which clauses of all lengths are allowed.

    We show, in fact, that there is no polynomial-time $\ratiofive$-approximation algorithm for MAX 5-SAT, and that there is no polynomial-time $\rationine$-approximation algorithm for MAX 9-SAT, again assuming UGC and $\mathsf{P} \neq \mathsf{NP}$. 
\end{abstract}

\paragraph{\footnotesize The Unique Games Conjecture} {\footnotesize A proof of the Unique Games Conjecture was recently announced by \mbox{OpenAI} \cite{OAI:UGC}. To the best of our knowledge it has not been independently verified yet. Thus, we continue to document which of our results depend on UGC. A proof of the Unique Games Conjecture would make our results stronger by only needing to depend on $\mathsf{P} \neq \mathsf{NP}$.}

\paragraph{\footnotesize AI Disclosure} {\footnotesize ChatGPT 6 Astra was used to obtain the results contained in the paper. The authors take full responsibility over the content in this paper. See Section~\ref{sec:disclosure} for a more detailed disclosure.}

\pagebreak

\section{Introduction}

The seminal works of Cook~\cite{Cook71} and Levin~\cite{Levin73} showed that Satisfiability (SAT) is \textsf{NP}-complete. In particular, there is no polynomial-time algorithm for solving SAT, unless \textsf{P}=\textsf{NP}. A lot of effort was therefore devoted, over the past five decades, to understanding the \emph{approximability} of SAT, or more precisely to the MAX SAT problem, where the goal is to find an assignment maximizing the total weight of satisfied clauses.

The PCP theorem~\cite{ALMSS98,AS98,D07} implies that there exists an absolute constant $\gamma > 0$ such that obtaining a polynomial-time \footnote{All approximation algorithms considered in this paper are assumed to run in polynomial time.} $(1-\gamma)$-approximation algorithm for MAX SAT is \textsf{NP}-hard. In other words, MAX SAT does not admit a Polynomial Time Approximation Scheme (PTAS), unless \textsf{P}=\textsf{NP}. H{\aa}stad~\cite{H01} went much further and showed that a $(\frac78+\eps)$-approximation of MAX SAT is \textsf{NP}-hard for all $\eps > 0$. Interestingly, the result of H{\aa}stad~\cite{H01} applies also to satisfiable instances of MAX $\{3\}$-SAT, the version of the MAX SAT problem in which all clauses are of size exactly 3. (In the general MAX SAT problem, clauses of arbitrary length are allowed.) H{\aa}stad's~\cite{H01} result is clearly tight for MAX $\{3\}$-SAT, as a random assignment satisfies every clause of length at least $3$ with probability at least~$\frac78$.

Raghavendra~\cite{R08} (see also \cite{RS09} and \cite{BCR15}) showed that, assuming UGC~\cite{khot02} and $\textsf{P}\ne\textsf{NP}$, the optimal approximation ratio for any MAX CSP, including in particular MAX SAT \footnote{Some care is needed when applying Raghavendra's result to MAX SAT, as the clauses of MAX SAT can grow as a function of the input size. See Appendix E of \cite{BHPZ25NAE} for an explanation as to why Raghavendra's theorem still applies.}, is equal to the integrality ratio of the \emph{basic} SDP relaxation of the problem. (The basic SDP relaxation of MAX SAT is given in Section~\ref{S-SDP}.) Furthermore, an approximation algorithm coming arbitrary close to this ratio can be obtained by a generic algorithm that rounds a solution to this relaxation. Unfortunately, Raghavendra's result does not specify what this integrality ratio is, and the generic rounding algorithm is far from practical as it amounts to enumerating, up to a certain precision, all rounding procedures from a fairly large family.

In view of H{\aa}stad's~\cite{H01} result, it was natural to ask whether there is a $\frac78$-approximation algorithm for the MAX SAT problem, with clauses of all sizes allowed. Karloff and Zwick~\cite{KZ97} and Zwick~\cite{zwick02}, showed that this is possible for MAX 3-SAT, the version of the problem in which clauses of length at most~3 are allowed. As mentioned above, a ratio of $\frac78$ is trivial if only clauses of length at least~$3$ are allowed. Our main result is that there is no $\frac78$-approximation algorithm for MAX SAT, when clauses of all lengths are allowed, assuming UGC and $\textsf{P}\ne \textsf{NP}$. In fact, we show that this is impossible even when all clause lengths are in $\{1,3,4,5\}$. \footnote{In the next version of the paper we will also show that this is impossible even if all clause lengths are in $\{1,4\}$. The proof of this claim is more complicated and we are working now on polishing it.}

\subsection{Variants of the MAX SAT problem}

MAX SAT refers to the version of the problem in which clauses of all lengths are allowed. If $K$ is a set of natural numbers, we let MAX $K$-SAT be the version of the problem in which only clauses whose lengths belong to~$K$ are allowed. For example MAX $\{1,4\}$-SAT allows only clauses of length 1 or 4, and MAX $\{3\}$-SAT allows only clauses of size~$3$. We also use MAX $k$-SAT as a short hand for MAX $\{1,2,\ldots,k\}$-SAT, i.e., when clauses of length up to~$k$ are allowed.

\subsection{Approximation algorithms for MAX SAT}

The MAX SAT problem was first formalized by Karp~\cite{Karp72}. The first nontrivial approximation algorithm was given by Johnson~\cite{J74} which had an approximation ratio of $\frac12$. After this, a $\frac34$-approximation algorithm was obtained by Yannakakis \cite{Y94} using flow techniques, and soon afterwards by Goemans and Williamson using LP rounding \cite{GW94}. Many other $\frac34$-approximation algorithms followed since. Obtaining approximation ratios larger than $\frac34$ seems to require semidefinite programming. This was first done by Goemans and Williamson \cite{GW95}. Avidor, Berkovitch, and Zwick \cite{ABZ05} obtained an algorithm with a ratio of $0.7968$ and a second algorithm with a \emph{conjectured}, not rigorously verified, ratio of $0.8434$. These are currently the best rigorously verified and conjectured ratios. However, there have been many other algorithms for MAX SAT in the literature, see \cite{LS81,KK89,GW94,FG95,AOH96,AHOH97,Asa97,AE98,CFZ99,Zwick99a,HO00,Trevisan00,AW02,Eng04,ZYH04,A06,CST11,PS11,Pol11,vZ12,PWZ14,BFNS15,PSWZ17}. For a survey of some related results, see Makarychev and Makarychev \cite{MM17}. 

It is known that the optimal approximation ratio of MAX 2-SAT, assuming UGC and $\textsf{P}\ne\textsf{NP}$, is about $0.94016567$ (see \cite{LLZ02}, \cite{Austrin07} and \cite{BHZ24}), and that the optimal approximation ratio of MAX 3-SAT is~$\frac78$ (see \cite{H01}, \cite{KZ97}, \cite{zwick02}). Halperin and Zwick \cite{HZ01} showed that the approximation ratio of MAX $\{1,4\}$-SAT is at most $0.8756$ \footnote{Halperin and Zwick \cite{HZ01} actually claimed an upper bound of $0.8754$, but misidentified the global maximum of a function used in their analysis. A slight modification of their construction, described in Section~\ref{S-HZ}, establishes an upper bound of $0.8756$.}, assuming UGC, and gave an approximation algorithm for MAX 4-SAT with a conjectured approximation ratio of $0.8721$. For results for the related MAX NAE-SAT problem, in which each clause is required to have both a true and a false literal, see \cite{BHPZ25NAE} and the references therein. 

\subsection{Our results}

\begin{table}[t]
\centering
\begin{tabular}{c@{\qquad}c@{\qquad}c@{\qquad}c@{\qquad}c}
\toprule
$k$ & Families & Clause types & Clause lengths $K$
    & Upper bound on $\alpha_K$ \\
\midrule
5 & 2 & 4 & $\{1,3,4,5\}$     & $0.87418758$ \\
6 & 3 & 7 & $\{1,3,4,5,6\}$   & $0.87401443$ \\
7 & 3 & 7 & $\{1,3,4,5,6,7\}$ & $0.87397813$ \\
8 & 3 & 7 & $\{1,3,4,5,8\}$   & $0.87354080$ \\
9 & 3 & 7 & $\{1,3,4,5,9\}$   & $0.87337371$ \\
\bottomrule
\end{tabular}
\caption{Upper bounds on the integrality ratios of the basic SDP relaxations of MAX $k$-SAT, for $5\le k\le 9$. The second and third columns give the number of variable families and the number of clause types used in the upper bound constructions. (These notions are explained in Sections \ref{S-SDP} and~\ref{S-HZ}.)}
\label{T-summary}
\end{table}

We show that there is no $\frac78$-approximation algorithm for MAX SAT, when clauses of all sizes are allowed, assuming UGC and $\mathsf{P} \neq \mathsf{NP}$. We obtain the following stronger results from which the $\frac78$ result follows as a corollary.

\begin{theorem}\label{thm:main}
    (a) There is no polynomial-time approximation algorithm for MAX $5$-SAT with an approximation ratio of at least $0.8742$, assuming UGC and $\mathsf{P} \neq \mathsf{NP}$. (b) There is no polynomial-time approximation algorithm for MAX $9$-SAT with an approximation ratio of at least $0.8734$, assuming UGC and $\mathsf{P} \neq \mathsf{NP}$.
\end{theorem}

\begin{corollary}
    There is no polynomial-time $\frac78$-approximation algorithm for \mbox{MAX SAT}, assuming~UGC.
\end{corollary}

We obtain a few additional results, summarized in Table~\ref{T-summary}.

In the next version of this paper we will also show that there is no polynomial-time approximation algorithm for MAX 4-SAT with an approximation ratio of at least $\frac78 - \gamma$ for some explicit $\gamma > 0$.

\subsection{Organization of the paper}

In Section~\ref{S-SDP} we describe the basic SDP relaxation of the MAX $k$-SAT problem. In Section~\ref{S-HZ} we give a concise description of the construction of Halperin and Zwick \cite{HZ01} for MAX 4-SAT. All our constructions may be seen as extensions of this construction. In Section~\ref{S-MAX-5-SAT} we give our new construction for MAX 5-SAT which is the first construction to give an upper bound below $\frac78$ for any version of MAX SAT. In Section~\ref{S-MAX-6-SAT} we discuss similar constructions for MAX $k$-SAT for $k=6,7,8,9$, including our construction for $k=9$ that yields our best upper bound for MAX SAT. 

\section{The basic SDP relaxation of MAX $k$-SAT}\label{S-SDP}

\begin{figure}[t]
\[
\boxed{
\begin{aligned}
\max\quad&
\sum_{C\in\cC}w_C\,\mu_C(\Omega_C^+)
\\[4pt]
\text{s.t.}\quad&
\bv_x\cdot\bv_x \EQ 1 \;,
&&x\in V\cup\{0\}
\\[2pt]
&\mu_C\in\Delta(\Omega_C)\;,
&&C\in\cC
\\[2pt]
&\Pr_{\alpha\sim\mu_C}[\alpha(x)=1]
\EQ\frac{1-\bv_0\cdot\bv_x}{2}\;,
&&x\in V(C) \;,\; C\in\cC
\\[5pt]
&\Pr_{\alpha\sim\mu_C}[\alpha(x)\ne\alpha(y)]
\EQ\frac{1-\bv_x\cdot\bv_y}{2}\;,
&&x,y\in V(C) \;,\; C\in\cC
\end{aligned}
}
\]
\caption{The basic SDP relaxation of MAX SAT.}\label{F-SDP}
\end{figure}

The basic SDP relaxation of the MAX SAT problem is given in Figure~\ref{F-SDP}, which is similar to the presentation of Raghavendra~\cite{R08}. A MAX SAT instance is specified in the following way. We let~$V$ be the set of variables and $\cC$ be the set of clauses. Each clause $C\in \cC$ is a finite collection of distinct literals, where each literal is either a variable $x\in V$ or its negation $\neg x$. A clause is not allowed to contain both $x$ and $\neg x$. Each clause has a nonnegative weight $w_C$ associated with it. We let $V(C)$ be the set of variables $x$ for which either $x$ or $\neg x$ appear in~$C$. We let $\Omega_C=\{\alpha:V(C)\to \{0,1\}\}$ be the set of all Boolean assignments to the variables of~$V(C)$.  We let~$\Omega^+_C$ be the set of all such assignments that satisfy~$C$, i.e., assign the value 1 to at least one literal in~$C$. (We let $\alpha(\neg x)=\neg \alpha(x)$, for every $x\in V(C)$.) Note that $|\Omega_C|=2^{|V(C)|}$ while $|\Omega^+_C|=2^{|V(C)|}-1$. We denote by $\mu_C\in \Delta(\Omega_C)$ a probability distribution over $\Omega_C$. More formally $\mu_C:\Omega_C\to[0,1]$ where $\sum_{\alpha\in \Omega_C}\mu_C(\alpha)=1$.

The relaxation has a unit vector $\bv_0$ representing~$0$, and a unit vector $\bv_x$ for each variable $x\in V$. For each clause $C\in\cC$ it has local distribution $\mu_C$ on the possible Boolean assignments to $\Omega_C$. We let
\[ \mu_C(\Omega^+_C) \EQ \sum_{\alpha\in\Omega^+_C}\mu_C(\alpha) \;,\]
be the probability $C$ is satisfied by a random assignment to its variables chosen according to the distribution~$\mu_C$. The objective is to maximize $\sum_{C\in\cC}w_C\,\mu_C(\Omega_C^+)$, the expected total weight of the clauses satisfied by their local distributions, under the constraints that the local distributions are consistent with inner products of the unit vectors, as specified by the relaxation. More specifically, we require
\[ \Pr_{\alpha\sim\mu_C}[\alpha(x)=1]
\EQ\frac{1-\bv_0\cdot\bv_x}{2}\quad,\quad
\Pr_{\alpha\sim\mu_C}[\alpha(x)\ne\alpha(y)]
\EQ\frac{1-\bv_x\cdot\bv_y}{2}\quad,\quad x,y\in V(C)\;,\]
where $\Pr_{\alpha\sim\mu_C}[\alpha(x)=1]$ denotes the probability that $x$ is assigned the value~$1$ by an assignment chosen randomly according to $\mu_C$. More precisely, 
\[\Pr_{\alpha\sim\mu_C}[\alpha(x)=1] \EQ \sum_{\substack{\alpha\in \Omega_C\\\alpha(x)=1}}\mu_C(\alpha) \quad,\quad
\Pr_{\alpha\sim\mu_C}[\alpha(x)\ne\alpha(y)] \EQ\sum_{\substack{\alpha\in \Omega_C\\\alpha(x)\ne\alpha(y)}}\mu_C(\alpha) \;.\]

It is easy to see that this is a relaxation of the problem. If $A:V\to\{0,1\}$ is a global assignment of Boolean values to the variables, we can let $\bv_x=\bv_0$, if $A(x)=0$, and $\bv_x=-\bv_0$, if $A(x)=1$. We define each local distribution $\mu_C$ to give probability~$1$ to the restriction of~$A$ to~$V(C)$. It is easy to see that the value of this solution is equal to the total weight of the clauses satisfied by~$A$.

Strictly speaking, the relaxation in Figure~\ref{F-SDP} is a vector program, not an SDP. However, it is easily converted into a corresponding SDP by letting $X_{xy}=\bv_x\cdot\bv_y$ and requiring $X\succeq 0$.

The relaxation in Figure~\ref{F-SDP} is equivalent to the relaxations given by Raghavendra \cite{R08} for general CSPs, specialized to the MAX SAT problem.

For MAX 2-SAT, MAX 3-SAT, and MAX 4-SAT, it is possible to obtain simplified versions of the relaxation that do not explicitly involve local distributions (see, e.g., \cite{BHZ24}, \cite{KZ97} and \cite{HZ01}). In principle this can be done for any clause size, but the number of constraints grows exponentially with~$k$ (e.g., \cite{KTW13}). Note that in the formulation given, the number of variables grows exponentially with~$k$.

If $I$ is a MAX SAT instance, we let $\OPT(I)$ be the total weight satisfied by an optimal Boolean assignment. We let $\SDP(I)$ be the value of an optimal solution of the relaxation. As explained, we have $\OPT(I)\le \SDP(I)$. For a given set $K$ of clause lengths, the integrality gap $\alpha_K$ of the relaxation on MAX $K$-SAT is defined as follows:
\[ \alpha_K \EQ \inf_I \, \frac{\OPT(I)}{\SDP(I)} \;,\]
where the infimum is over MAX $K$-SAT instances. We also let
\[ \alpha_k \EQ \alpha_{\{1,2,\ldots,k\}} \;.\]

\section{The Halperin-Zwick construction for MAX $\{1,4\}$-SAT}\label{S-HZ}

Halperin and Zwick \cite{HZ01} described a simple construction that established an upper bound of $\alpha_4\le\alpha_{\{1,4\}}\le 0.8756$, coming just short of proving that a polynomial-time $\frac78$-approximation algorithm for MAX 4-SAT is impossible. We begin with a concise description of a slightly modified version of their construction as all our improved constructions may be viewed as modifications and extensions of this basic construction. We also correct a numerical error in \cite{HZ01}.

\begin{table}[t]
\[
\begin{array}{c}
\begin{array}{c@{\qquad}c@{\qquad}c}
 \toprule
 \text{Type}&\text{Clause}&\text{Total weight}\\
 \midrule
 1&\neg{X}& w_1=17 \\
 2&X_1\vee X_2\vee X_3\vee\neg{Y}& w_2=30 \\
 3&Y_1\vee Y_2\vee Y_3\vee Y_4& w_3=12 \\
 \bottomrule
\end{array} \\ \\
\text{(a) Variable families and clause types.}
\end{array}
\hspace*{0.5cm}
\begin{array}{c}
\begin{array}{c}
C \EQ
\begin{pmatrix}
1&a&b\\
a & \frac{1+2b}{3} & \frac{b+3a-1}{3} \\
b & \frac{b+3a-1}{3} & \frac{1-2b}{3}
\end{pmatrix}\\[25pt]
a \EQ \frac{4+\sqrt{17}}{15}\approx0.541540 \;,\; b\EQ \frac15
\end{array} \\ \\
\text{(b) Inner products.}
\end{array}
\]

\vspace*{10pt}
\centering
\setlength{\tabcolsep}{3pt}

\begin{tabular}[t]{@{}cc@{}}
\multicolumn{2}{c}{$\boldsymbol{\neg X}$}\\[2mm]
\toprule
$X$ & Probability \\
\midrule
0 & $\frac{1+a}{2}$ \\[4pt]
1 & $\frac{1-a}{2}$ \\
\bottomrule
\end{tabular}
\hspace{1cm}
\begin{tabular}[t]{@{}ccccc@{}}
\multicolumn{5}{c}{$\boldsymbol{X_1\vee X_2\vee X_3\vee\neg Y}$}\\[2mm]
\toprule
$X_1$ & $X_2$ & $X_3$ & $Y$ & Probability \\
\midrule
1 & 1 & 1 & 0 & $\frac{2+b-3a}{6}$ \\[4pt]
1 & 0 & 0 & 1 & $\frac{1-b}{6}$ \\[4pt]
0 & 1 & 0 & 1 & $\frac{1-b}{6}$ \\[4pt]
0 & 0 & 1 & 1 & $\frac{1-b}{6}$ \\[4pt]
0 & 0 & 0 & 0 & $\frac{1+2b+3a}{6}$ \\
\bottomrule
\end{tabular}
\hspace{1cm}
\begin{tabular}[t]{@{}ccccc@{}}
\multicolumn{5}{c}{$\boldsymbol{Y_1\vee Y_2\vee Y_3\vee Y_4}$}\\[2mm]
\toprule
$Y_1$ & $Y_2$ & $Y_3$ & $Y_4$ & Probability \\
\midrule
1 & 0 & 0 & 0 & $\frac{1+b}{6}$ \\[4pt]
0 & 1 & 0 & 0 & $\frac{1+b}{6}$ \\[4pt]
0 & 0 & 1 & 0 & $\frac{1+b}{6}$ \\[4pt]
0 & 0 & 0 & 1 & $\frac{1+b}{6}$ \\[4pt]
1 & 1 & 1 & 1 & $\frac{1-2b}{3}$ \\
\bottomrule
\end{tabular}

\vspace*{15pt}
(c) Local distributions.
\[\begin{array}{c}
x^*\EQ 0.3444012451\ldots\\
y^*\EQ 0.4394709493\ldots\\[10pt]
\text{(d) Optimal fractions}
\end{array}
\hspace*{1.5cm}
\begin{array}{c}
\alpha_{\{1,4\}} \LE 0.8755500085\\[10pt]
\text{(e) Upper bound}
\end{array}
\]
\caption{The Halperin-Zwick \cite{HZ01} construction.}\label{T-HZ}
\end{table}

A compact description of the construction appears in Table~\ref{T-HZ}. 
For every $n\ge 4$, we construct an instance of MAX $\{1,4\}$-SAT on the $2n$ variables $X=\{x_1,x_2,\ldots,x_n\}$ and $Y=\{y_1,y_2,\ldots,y_n\}$. We refer to~$X$ and~$Y$ as \emph{families} of variables. The instance contains clauses from three different \emph{types}, shown in Table~\ref{T-HZ}(a). Each clause of the first type is of the form $\neg{X}$. Each clause of the second type is of the form $X_1\lor X_2 \lor X_3\lor \neg{Y}$. Each clause of the third type is of the form $Y_1\lor Y_2 \lor Y_3\lor Y_4$. When describing clause types, we use $X_i$ and $Y_i$ to denote arbitrary $x$ or $y$ variables. When two such variables appear in the same clause, we require them to be distinct. Thus, there are exactly~$n$ clauses of the first type, $\binom{n}{3}n$ clauses of the second type, and $\binom{n}{4}$ clauses of the third type. The total weight of all the clauses from the three types are 17, 30 and 12, respectively. These weights are distributed evenly among all the clauses of the given type. For example, each one of the $\binom{n}{4}$ clauses of the third type has individual weight ${12}/{\binom{n}{4}}$. This completely specifies an instance $I_n$ of MAX $\{1,4\}$-SAT for every $n\ge 4$. (The total weights 17, 30 and 12 were chosen for their simplicity. They can be slightly optimized. They can also be normalized to $\frac{17}{59},\frac{30}{59}$ and $\frac{12}{59}$, if desired.)

To obtain an upper bound on the integrality ratio of the MAX $\{1,4\}$-SAT relaxation, we need to obtain a lower bound on $\SDP(I_n)$ and an upper bound on $\OPT(I_n)$. To obtain a lower bound on $\SDP(I_n)$ we simply present a carefully chosen feasible solution of the relaxation.
We then bound $\OPT(I_n)$ by noticing that since the instance is symmetric, the total weight of the clauses satisfied by a given assignment depends only on the number of variables from each of the families~$X$ and~$Y$ that evaluate to true.

We begin by constructing a concrete solution of the relaxation. The solution is composed of $2n+1$ unit vectors $\bv_0,\bv_{x_1},\dots,\bv_{x_n},\bv_{y_1},\dots,\bv_{y_n}$ and a corresponding local distribution for each clause. The local distribution of a clause should be consistent with the inner products between the vectors corresponding to the variables of the clause.

We begin by choosing three vectors $\bv_0,\bv_X,\bv_Y$ with the inner products specified by the matrix~$C$ given in Table~\ref{T-HZ}(b). Note that $\bv_0$ is a unit vector while~$\bv_X$ and~$\bv_Y$ are not. In principle, for two variable families~$X$ and~$Y$, the inner product matrix~$C$ is specified by five independent parameters $c_{0X},c_{0Y},c_{XX},c_{XY},c_{YY}$. We explain later why we choose a matrix~$C$ that depends only on two free parameters $a=c_{0X}$ and $b=c_{0Y}$. (Note that we must have $-1\le b\le\frac{1}{3}$.) For simplicity we also choose the specific parameters $a = \frac{4+\sqrt{17}}{15}\approx0.541540$ and $b=\frac15$ as shown in the table.\footnote{We note that these parameters can be optimized to give a slightly better result.} Thus,
\[ c_{XX} \EQ \bv_X\cdot \bv_X \EQ \frac{1+2b}{3} \EQ \frac{7}{15} \quad,\quad c_{YY} \EQ \bv_Y\cdot \bv_Y \EQ \frac{1-2b}{3} \EQ \frac15 \;.\]
\[ c_{0X} \EQ \bv_0\cdot \bv_X \EQ a  \quad,\quad c_{0Y} \EQ \bv_0\cdot \bv_Y \EQ b \EQ \frac15 
\quad,\quad c_{XY} \EQ \bv_X\cdot \bv_Y \EQ \frac{b+3a-1}{3} \;.\]
To show that such three vectors $\bv_0,\bv_X,\bv_Y$ exist, we need to show that $C\succeq0$, i.e., that the matrix~$C$ is Positive Semidefinite (PSD). Since $-1< b< \frac13$, it is not difficult to check that 
\[
C\succeq0\quad\Longleftrightarrow\quad\det(C)\GE 0\;.
\]
In our case
\[
\det(C) \EQ 
\frac19\left(
(1+b)(1-3b)(1+2b-3a^2)
-(1-b)^2(3a-1)^2
\right) \;.
\]
For the choice $a = \frac{4+\sqrt{17}}{15}\approx0.541540$ and $b=\frac15$ we have $\det(C)=0$. In fact, this is how~$a$ was set given the choice $b=\frac15$. For the simple choice of~$a$ and~$b$, we can actually construct the vectors $\bv_0,\bv_X$ and $\bv_Y$ directly:
\[
\bv_0 \EQ (1,0) \quad,\quad
\bv_X \EQ \left(a, 2a-\frac23\right) \quad,\quad
\bv_Y \EQ (b,2b)\;.
\]
A simple calculation then shows that the inner products are indeed those specified by~$C$. 

We now choose unit vectors $\bu_{x_1},\dots,\bu_{x_n},\bu_{y_1},\dots,\bu_{y_n}$ that are orthogonal to each other and to $\bv_0,\bv_X,\bv_Y$. (This can be easily done by adding $2n$ new coordinates.) The vector $\bv_{x_i}$ and $\bv_{y_j}$ are now defined as follows:
\[
\bv_{x_i} \Eq \bv_X+\sqrt{1-c_{XX}}\,\bu_{x_i}
\Eq \bv_X+\sqrt{\frac{2(1-b)}{3}}\,\bu_{x_i}
\;,\;
\bv_{y_j} \Eq \bv_Y+\sqrt{1-c_{YY}}\,\bu_{y_j}
\Eq \bv_Y+\sqrt{\frac{2(1+b)}{3}}\,\bu_{y_j}
\;.
\]
It is easy to check that $\bv_{x_1},\dots,\bv_{x_n},\bv_{y_1},\dots,\bv_{y_n}$ are all unit vectors and that 
\[ \bv_{0}\cdot \bv_{x_i} \Eq c_{0X} \;,\; \bv_{0}\cdot \bv_{y_j} \Eq c_{0Y}\;,
\bv_{x_i}\cdot \bv_{x_j} \Eq c_{XX} \;,\; \bv_{x_i}\cdot \bv_{y_j} \Eq c_{XY} \;,\; \bv_{y_i}\cdot \bv_{y_j} \Eq c_{YY}\;,\]
where the third and fifth relations %
assume that $i\ne j$.

To complete the solution of the SDP relaxation, we need to specify a local distribution for each clause which is consistent with the corresponding inner products. These local distributions are given in Table~\ref{T-HZ}(c). An interesting property of these local distributions is that they satisfy all the 4-clauses with probability~$1$. In fact, all local assignments with positive probabilities either satisfy all 4 literals in such clauses, or exactly one literal. Such a choice is consistent with the inner product matrix~$C$ only if~$C$ has the form shown in Table~\ref{T-HZ}(b). (The choice $b=\frac15$, is not dictated.)

We need to verify that the local distributions given in Table~\ref{T-HZ}(c) are consistent with the inner products specified by $C$. More specifically, we need to verify that
\[\begin{array}{c} 
\ds \Pr[X=1] \EQ \frac{1-c_{0X}}{2} \EQ \frac{1-a}{2} \quad,\quad \Pr[Y=1] \EQ \frac{1-c_{0Y}}{2} \EQ\frac{1-b}{2} \;, \\[10pt]
\ds \Pr[X_1\ne X_2] \EQ \frac{1-c_{XX}}2 \EQ\frac{1-b}{3} \quad,\quad
\Pr[Y_1\ne Y_2] \EQ \frac{1-c_{YY}}2 \EQ\frac{1+b}{3} \;,\\[10pt]
\ds \Pr[X\ne Y] \EQ \frac{1-c_{XY}}2=\frac{4-b-3a}{6} \;,
\end{array}\]
where $X,X_1,X_2$ and $Y,Y_1,Y_2$ stand for any variables from the~$X$ and~$Y$ families, with $X_1,X_2$ and $Y_1,Y_2$ corresponding to distinct variables from the respective families. These equalities should hold in every local distribution containing the relevant variables. The verification of the equalities is straightforward.

For the first distribution, we immediately have
\[
\Pr[X=1]\EQ\frac{1-a}{2}\;.
\]
For the second distribution, we have
\[
\begin{aligned}
\Pr[X_i=1]
&\EQ \frac{2+b-3a}{6}+\frac{1-b}{6} \EQ \frac{1-a}{2}\;,\\[4pt]
\Pr[Y=1]
&\EQ 3\cdot\frac{1-b}{6}\EQ\frac{1-b}{2}\;,\\[4pt]
\Pr[X_i\ne X_j]
&\EQ 2\cdot\frac{1-b}{6}\EQ\frac{1-b}{3}\;,\\[4pt]
\Pr[X_i\ne Y]
&\EQ\frac{2+b-3a}{6}+2\cdot\frac{1-b}{6} \EQ\frac{4-b-3a}{6} \;.
\end{aligned}
\]
Finally, for the third distribution, we have
\[
\begin{aligned}
\Pr[Y_i=1] &\EQ \frac{1+b}{6}+\frac{1-2b}{3} \EQ \frac{1-b}{2}\;,\\[4pt]
\Pr[Y_i\ne Y_j] &\EQ 2\cdot\frac{1+b}{6} \EQ \frac{1+b}{3}\;.
\end{aligned}
\]
This completes the verification.

Since the local distributions satisfy each 1-clause with probability $\frac{1+a}{2}$, and every 4-clause with probability~$1$, we get that
\[\mathrm{SDP}(I_n) \GE w_1\,\frac{1+a}{2} + w_2 + w_3\;.\]

As observed above, to determine the total weight of the clauses satisfied by a given Boolean assignment to the variables of $I_n$ we only need to specify how many of the~$X$ variables, and how many of the~$Y$ variables are set to~$1$. Let~$x$ and~$y$ be the fraction of the~$X$ and~$Y$ variables set to~$1$. As $n\to\infty$, the total weight of the clauses satisfied by such an assignment approaches
\begin{equation}\label{eq:hz-integral}
  F_4(x,y)=
  w_1(1-x)+w_2\bigl(1-(1-x)^3y\bigr)
       +w_3\bigl(1-(1-y)^4\bigr)\;.
\end{equation}
Thus, the upper bound on the integrality ratio of the relaxation approaches 
\begin{align}\max_{0\le x,y\le 1} \frac{\ds  w_1(1-x)+w_2\bigl(1-(1-x)^3y\bigr)
       +w_3\bigl(1-(1-y)^4\bigr)}{\ds w_1\,\frac{1+a}{2} + w_2 + w_3} \;.\label{eq:hz-ratio}\end{align}
The only thing left to do, therefore, is to optimize over~$x$ and~$y$. It is not difficult to check, by differentiating, that for every fixed $z=1-x$, the optimal choice of~$y$ is \[y^*=\max\left\{0, 1-\left(\frac{w_2}{4w_3}\right)^{1/3}z\right\}\;.\]
When $w_2\le 4w_3$, as in our case, we have $y^*= 1-\left(\frac{5}{8}\right)^{1/3}z$. Plugging this optimal value of~$y$, we get that we need to maximize the following univariate polynomial:
\[ F(z) \EQ w_2+w_3+w_1z-w_2z^3
+\frac34\left(\frac{w_2^4}{4w_3}\right)^{1/3}z^4\;.\]
For our choice of $w_1=17$, $w_2=30$ and $w_3=12$, we get that the optimal settings of $x$, $y$ are:
\[
\begin{aligned}
x^*&\EQ 1-z^* \EQ 0.3444012451\ldots,\\
y^*&\EQ 1-\left(\frac{5}{8}\right)^{1/3}z^* \EQ 0.4394709493\ldots,
\end{aligned}
\]
and the resulting upper bound on the integrality ratio is
\[ \alpha_{\{1,4\}} \LE 0.8755500085 \;.\]
This upper bound can be improved to $0.8755341563$ by optimizing over $w_1,w_2,w_3$ and over $a$ and $b$, but the parameters then are not as clean as the ones used here. Halperin and Zwick \cite{HZ01} claim an upper bound of $0.8754$, using essentially the same construction presented here, but unfortunately they incorrectly optimized (\ref{eq:hz-ratio}), leading to a slight numerical error.

\section{A new construction for MAX $\{1,3,4,5\}$-SAT}\label{S-MAX-5-SAT}

The new construction for MAX $\{1,3,4,5\}$-SAT is given in
Table~\ref{T-five}. It has many similarities with the
MAX $\{1,4\}$-SAT construction given in the previous section.
We again use only two families of variables, $X$ and $Y$,
each containing $n$ variables.
This time, however, we have four clause types, of sizes
$1,3,4$ and $5$, respectively, obtained by replacing the
positive 4-clauses of the previous construction by positive
5-clauses and introducing an additional type of 3-clauses.
There are $n$, $\binom{n}{2}(n-2)$, $\binom{n}{3}n$
and $\binom{n}{5}$ clauses of the four types, respectively.
Their total weights are $32$, $19$, $57$ and $26$,
respectively, distributed evenly among the clauses of each
type. This completely specifies an instance $I_n$ of
MAX $\{1,3,4,5\}$-SAT for every $n\ge5$.

\begin{table}[!t]
\centering
\setlength{\tabcolsep}{3pt}
\setlength{\arraycolsep}{3pt}
\[\begin{array}{c}
\begin{array}{c@{\quad}c@{\quad}c}
\toprule
\text{Type}&\text{Clause}&\text{Total weight}\\
\midrule
1&\neg X& w_1=32 \\
2&Y_1\vee Y_2\vee\neg Y_3& w_2=19 \\
3&X_1\vee X_2\vee X_3\vee\neg Y& w_3=57 \\
4&Y_1\vee Y_2\vee Y_3\vee Y_4\vee Y_5& w_4=26 \\
\bottomrule
\end{array}\\ \\[-4pt]
\text{(a) Variable families and clause types.}
\end{array}
\hspace*{0.25cm}
\begin{array}{c}
C \EQ \begin{pmatrix}
1&a&\frac13\\
a&\frac59&a-\frac29\\
\frac13&a-\frac29&\frac13
\end{pmatrix}\\[25pt]
a \EQ \frac{2+\sqrt{13}}9\approx0.622839\\[15pt]
\text{(b) Inner products.}
\end{array}\]
\vspace{20pt}
\begin{tabular}[t]{@{}cc@{}}
\multicolumn{2}{c}{$\boldsymbol{\neg X}$}\\[2mm]
\toprule
$X$ & Prob \\
\midrule
0 & $\frac{1+a}{2}$ \\[4pt]
1 & $\frac{1-a}{2}$ \\
\bottomrule
\end{tabular}
\hspace{0.65cm}
\begin{tabular}[t]{@{}cccc@{}}
\multicolumn{4}{c}{$\boldsymbol{Y_1\vee Y_2\vee\neg Y_3}$}\\[2mm]
\toprule
$Y_1$ & $Y_2$ & $Y_3$ & Prob \\
\midrule
1 & 1 & 0 & $\frac{1}{6}$ \\[4pt]
1 & 0 & 1 & $\frac{1}{6}$ \\[4pt]
0 & 1 & 1 & $\frac{1}{6}$ \\[4pt]
0 & 0 & 0 & $\frac{1}{2}$ \\
\bottomrule
\end{tabular}
\hspace{0.65cm}
\begin{tabular}[t]{@{}ccccc@{}}
\multicolumn{5}{c}{$\boldsymbol{X_1\vee X_2\vee X_3\vee\neg Y}$}\\[2mm]
\toprule
$X_1$ & $X_2$ & $X_3$ & $Y$ & Prob \\
\midrule
1 & 1 & 1 & 0 & $\frac{7}{18}-\frac a2$ \\[4pt]
1 & 0 & 0 & 1 & $\frac{1}{9}$ \\[4pt]
0 & 1 & 0 & 1 & $\frac{1}{9}$ \\[4pt]
0 & 0 & 1 & 1 & $\frac{1}{9}$ \\[4pt]
0 & 0 & 0 & 0 & $\frac{5}{18}+\frac a2$ \\
\bottomrule
\end{tabular}
\hspace{0.65cm}
\begin{tabular}[t]{@{}cccccc@{}}
\multicolumn{6}{c}{$\boldsymbol{Y_1\vee Y_2\vee Y_3\vee Y_4\vee Y_5}$}\\[2mm]
\toprule
$Y_1$ & $Y_2$ & $Y_3$ & $Y_4$ & $Y_5$ & Prob \\
\midrule
1 & 0 & 0 & 0 & 0 & $\frac16$ \\[4pt]
0 & 1 & 0 & 0 & 0 & $\frac16$ \\[4pt]
0 & 0 & 1 & 0 & 0 & $\frac16$ \\[4pt]
0 & 0 & 0 & 1 & 0 & $\frac16$ \\[4pt]
0 & 0 & 0 & 0 & 1 & $\frac16$ \\[4pt]
1 & 1 & 1 & 1 & 1 & $\frac16$ \\
\bottomrule
\end{tabular}
\par
\vspace*{-10pt}
(c) Local distributions.
\[
\begin{array}{c}
x_*\approx 0.373856370162\ldots\\
y_*\approx 0.477315936547\ldots\\[10pt]
\text{(d) Optimal fractions.}
\end{array}
\hspace*{1cm}
\begin{array}{c}
\alpha_{\{1,3,4,5\}} \LE 0.8741875750\\[10pt]
\text{(e) Upper bound.}
\end{array}
\]
\caption{Two-family construction for MAX $\{1,3,4,5\}$-SAT.}\label{T-five}
\end{table}

As in the previous construction, we choose three vectors
$\bv_0,\bv_X,\bv_Y$ with the inner products specified by
the matrix~$C$ given in Table~\ref{T-five}(b).
The matrix~$C$ now depends on only one free parameter,
$a=c_{0X}$. We explain below why this is the case.

As $c_{YY}-c_{0Y}^2=\frac29>0$, we again have
\[
C\succeq0\quad\Longleftrightarrow\quad\det(C)\GE0\;.
\]
A simple calculation shows that
\[
\det(C) \EQ \frac{2}{27}(1+4a-9a^2)\;.
\]
We thus choose
\[
a \EQ \frac{2+\sqrt{13}}9 \approx 0.622839\;,
\]
which is the largest value of~$a$ for which $C\succeq0$.

As in the previous construction, we have $\det(C)=0$,
and we can construct the three vectors directly:
\[
\bv_0 \EQ (1,0)\quad,\quad
\bv_X \EQ \left(a,\sqrt{2}\left(a-\frac13\right)\right)
\quad,\quad
\bv_Y \EQ \left(\frac13,\frac{\sqrt{2}}3\right)\;.
\]
A simple calculation shows that their inner products
are indeed those specified by~$C$.

We obtain the unit vectors $\bv_{x_i}$ and $\bv_{y_j}$
as in the previous section:
\[
\bv_{x_i} \EQ \bv_X+\frac23\bu_{x_i}
\quad,\quad
\bv_{y_j} \EQ \bv_Y+\sqrt{\frac23}\bu_{y_j}\;,
\]
where the unit vectors $\bu_{x_i}$ and $\bu_{y_j}$ are
orthogonal to each other and to $\bv_0,\bv_X,\bv_Y$.

To complete the solution of the SDP relaxation, we need
to specify a local distribution for each clause which is
consistent with the corresponding inner products.
These local distributions are given in Table~\ref{T-five}(c).
As in the previous construction, all local assignments
with positive probability for non-unit clauses satisfy
either all the literals of the clause or exactly one literal.

We briefly explain why the matrix~$C$ now depends on only
one free parameter.
Let $a=c_{0X}$ and, temporarily, let $b=c_{0Y}$.
The local distribution for the mixed 4-clauses imposes,
as in the previous construction, the relations
\[
c_{XX} \EQ \frac{1+2b}{3}
\quad,\quad
c_{XY} \EQ \frac{b+3a-1}{3}\;.
\]
The chosen support of the local distribution for the
3-clauses imposes the relation $c_{YY}=b$, while the
chosen support of the local distribution for the
5-clauses imposes the relation $c_{YY}=\frac{1-b}{2}$.
We must therefore have
\[
b \EQ \frac{1-b}{2}\;,
\]
which gives $b=\frac13$.
Substituting this value, we obtain the matrix~$C$
given in Table~\ref{T-five}(b).

We need to check that the local distributions are
consistent with the inner products specified by~$C$,
i.e., that
\[
\Pr[X=1] \EQ \frac{1-a}{2}
\quad,\quad
\Pr[Y=1] \EQ \frac13\;,
\]
and
\[
\Pr[X_1\ne X_2] \EQ \frac29
\quad,\quad
\Pr[X\ne Y] \EQ \frac{11}{18}-\frac a2
\quad,\quad
\Pr[Y_1\ne Y_2] \EQ \frac13\;.
\]
Each equality should hold in every local distribution containing the relevant variable or pair of variables.

The verification is again straightforward. In the first distribution we have
\[
\Pr[X=1] \EQ \frac{1-a}{2}\;.
\]
In the second distribution we have
\[
\Pr[Y_i=1] \EQ 2\cdot\frac16 \EQ \frac13
\quad,\quad
\Pr[Y_i\ne Y_j] \EQ 2\cdot\frac16 \EQ \frac13\;.
\]
In the third distribution we have
\[
\Pr[X_i=1] \EQ \frac7{18}-\frac a2+\frac19
             \EQ \frac{1-a}{2}
\quad,\quad
\Pr[Y=1] \EQ 3\cdot\frac19 \EQ \frac13\;,
\]
\[
\Pr[X_i\ne X_j] \EQ 2\cdot\frac19 \EQ \frac29
\quad,\quad
\Pr[X_i\ne Y] \EQ \frac7{18}-\frac a2+2\cdot\frac19
               \EQ \frac{11}{18}-\frac a2\;.
\]
Finally, in the fourth distribution we have
\[
\Pr[Y_i=1] \EQ 2\cdot\frac16 \EQ \frac13
\quad,\quad
\Pr[Y_i\ne Y_j] \EQ 2\cdot\frac16 \EQ \frac13\;.
\]
In all the above calculations involving two variables
from the same family we assume that $i\ne j$.

Since the local distributions satisfy each unit clause
with probability $\frac{1+a}{2}$ and every non-unit clause
with probability~$1$, we get that
\[
\SDP(I_n) \GE w_1\,\frac{1+a}{2}+w_2+w_3+w_4\;.
\]

As in the previous construction, let $x$ and $y$ be the
fractions of the $X$ and $Y$ variables set to~$1$.
As $n\to\infty$, the total weight of the clauses satisfied
by such an assignment approaches
\[
F_5(x,y) \EQ 
w_1(1-x)+w_2\bigl(1-y(1-y)^2\bigr)
+w_3\bigl(1-(1-x)^3y\bigr)
+w_4\bigl(1-(1-y)^5\bigr)\;.
\]
The convergence is uniform over all assignments.
Thus, the construction gives the following upper bound
on the integrality ratio:
\[
\alpha_{\{1,3,4,5\}} \LE
\frac{\ds\max_{0\le x,y\le1}F_5(x,y)}
     {\ds w_1\,\frac{1+a}{2}+w_2+w_3+w_4}\;.
\]

Again, we can reduce the optimization problem to a
univariate optimization problem.
This time we optimize over $x$ for every fixed~$y$.
Let $z=1-x$. By differentiating with respect to~$z$,
we get
\[
\frac{\partial F_5}{\partial z}
\EQ w_1-3w_3yz^2\;.
\]
The optimal value of~$z$ for every fixed $y>0$ is thus
\[
z^*(y) \EQ
\min\left\{1,\sqrt{\frac{w_1}{3w_3y}}\right\}\;,
\]
with $z^*(0)=1$.

\begin{figure}[t]
\centering
\begin{tikzpicture}

\pgfmathsetmacro{\maxfivea}{(2+sqrt(13))/9}
\pgfmathsetmacro{\maxfiveD}{118+16*\maxfivea}
\pgfmathsetmacro{\ybreak}{32/171}

\begin{axis}[
    width=0.88\textwidth,
    height=0.52\textwidth,
    xmin=0, xmax=1,
    ymin=0.842, ymax=0.877,
    xlabel={$y$},
    ylabel={$G(y)/D$},
    xtick={0,0.2,0.4,0.6,0.8,1},
    ytick={0.845,0.850,0.855,0.860,0.865,0.870,0.875},
    scaled y ticks=false,
    yticklabel style={
        /pgf/number format/fixed,
        /pgf/number format/fixed zerofill,
        /pgf/number format/precision=3
    },
    tick label style={font=\small},
    label style={font=\small},
    legend style={
        font=\small,
        draw=none,
        fill=none,
        at={(0.97,0.04)},
        anchor=south east
    },
    samples=400,
    axis lines=box,
    clip=true,
]

\addplot[blue, very thick, domain=0:\ybreak]
    {(134 - 57*x - 19*x*(1-x)^2
      - 26*(1-x)^5)/\maxfiveD};
\addlegendentry{$G(y)/D$}

\addplot[blue, very thick, domain=\ybreak:1, forget plot]
    {(102 - 19*x*(1-x)^2 - 26*(1-x)^5
      + (64/3)*sqrt(32/(171*x)))/\maxfiveD};

\addplot[red, dashed, thick, domain=0:1]
    {7/8};
\addlegendentry{$7/8$}

\addplot[gray, densely dotted, forget plot]
    coordinates {
        (\ybreak,0.842)
        (\ybreak,0.877)
    };

\node[anchor=south west, font=\small]
    at (axis cs:\ybreak,0.8425)
    {$\frac{32}{171}$};

\addplot[blue, only marks, mark=*, mark size=2pt,
         forget plot]
    coordinates {
        (0.1577107552,0.8741589859)
        (0.2387682292,0.8722006047)
        (0.4773159365,0.8741875750)
    };

\node[anchor=north, font=\scriptsize, yshift=-4pt]
    at (axis cs:0.1577107552,0.8741589859)
    {$0.1577$};

\node[anchor=north, font=\scriptsize, yshift=-4pt]
    at (axis cs:0.2387682292,0.8722006047)
    {$0.2388$};

\node[anchor=north, font=\scriptsize, yshift=-4pt]
    at (axis cs:0.4773159365,0.8741875750)
    {$0.4773$};

\end{axis}
\end{tikzpicture}

\caption{The function $G(y)/D$ for the
MAX $\{1,3,4,5\}$-SAT construction, where
$D=118+16a$ is the value of the SDP solution.
The function has two local maxima and one local minimum.
Both local maxima are below $\frac78$.}
\label{F-max5-ratio}
\end{figure}
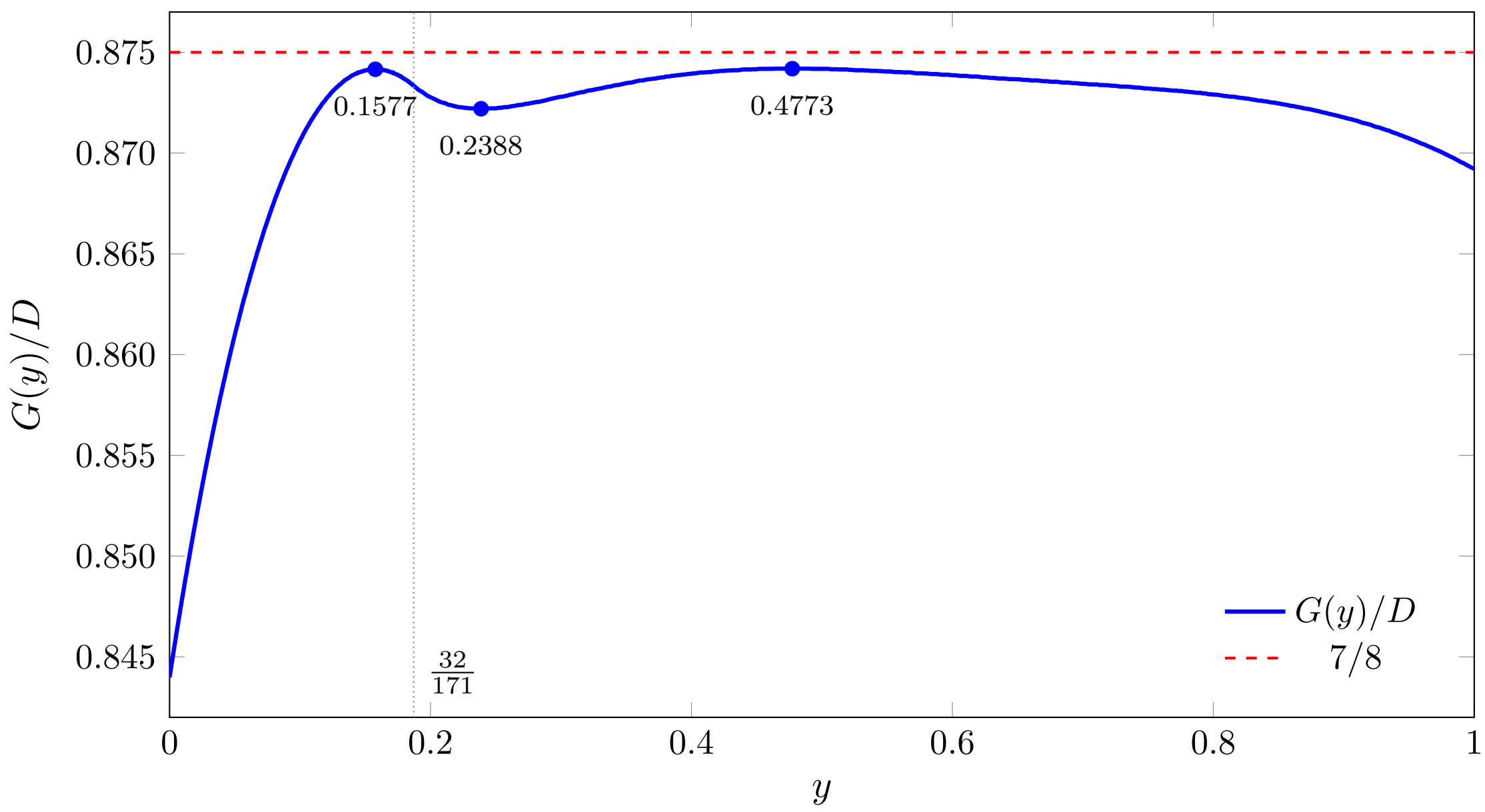

We are therefore left with the problem of maximizing
the following function over $0\le y\le1$:
\[
G(y) \EQ
w_2+w_3+w_4-w_2y(1-y)^2-w_4(1-y)^5
+\begin{cases}
w_1-w_3y \;,
& y\le\frac{w_1}{3w_3},\\[6pt]
\ds\frac{2w_1}{3}\sqrt{\frac{w_1}{3w_3y}} \;,
& y>\frac{w_1}{3w_3} \;.
\end{cases}
\]
A plot of the function $G(y)/D$, where $D=118+16a$ is the value of the exhibited SDP solution, is given in Figure~\ref{F-max5-ratio}. By differentiating, we find that it has two local maxima and one local minimum. The values attained by the two local maxima are almost identical. The local maximum at $y^*=0.477315936547\ldots$, which corresponds to $x^*= 1-z^*(y^*)=0.373856370162\ldots$, attains a slightly higher value of approximately $0.874187574995$. (The local maximum at $y^*=0.1577107552\ldots$ attains a value of $0.8741589859$.) We thus have:
\[
\alpha_5 \LE \alpha_{\{1,3,4,5\}}
\LT 0.8741875750 \LT \frac78\;.
\]
We can obtain a slightly better result by optimizing the weights. The optimized weights make the values attained by the two local maxima equal, but the improvement obtained is fairly minor (less than $10^{-5}$).

\section{Additional constructions for MAX $k$-SAT,
$6\le k\le 9$}\label{S-MAX-6-SAT}

We describe several extensions of the MAX $\{1,3,4,5\}$-SAT
construction of the previous section that give improved upper
bounds on the integrality ratios of MAX $k$-SAT, for $6\le k\le9$.
Complete descriptions of the constructions are given in
Tables~\ref{T-two6}--\ref{T-three9} of Appendix~\ref{A}. These constructions involve fairly large integer weights. We suppose that these weights can be substantially reduced with only a minor deterioration of the resulting upper bound, as was the case with the MAX 4-SAT (Table~\ref{T-HZ}) and MAX 5-SAT (Table~\ref{T-five}) constructions, but we did not pursue this direction.

To shorten the tables, we sometimes give only one representative of the assignments obtained from each other by permuting variables belonging to the same family and having the same sign in the clause. The multiplicity~$m$ indicates the number of assignments represented by a row, each having the displayed probability.

All these constructions retain the two families of variables
$X$ and $Y$, and the four clause types used in the construction
of the previous section, though the weights assigned to the
clause types are modified.
Furthermore, the inner products between the vectors
$\bv_0,\bv_X,\bv_Y$ are unchanged.
Thus, all the constructions use the same parameter
$a=(2+\sqrt{13})/9$.

\subsection{Constructions using two families of variables}
The simplest extensions of the construction of the previous
section, given in Tables~\ref{T-two6} to~\ref{T-two9}, are obtained by introducing just one additional clause
type, without introducing a new family of variables.
We use $X^pY^q$ to denote a clause type containing $p$ positive
literals from family~$X$ and $q$ positive literals from family~$Y$.
The additional clause types used in the four constructions,
together with the upper bounds obtained, are given in Table~\ref{T-add}.

\begin{table}[t]
\begin{center}
\begin{tabular}{c@{\qquad}c@{\qquad}c}
\toprule
$k$ & Additional clause type & Upper bound on $\alpha_k$ \\
\midrule
6 & $X^2Y^4$ & $0.8741802578$ \\
7 & $X^4Y^3$ & $0.8741313135$ \\
8 & $X^8$    & $0.87359549$ \\
9 & $X^9$    & $0.8734561883$ \\
\bottomrule
\end{tabular}
\end{center}
\caption{Additional clause types added to the MAX 5-SAT constructions to obtain the first improved constructions for MAX $k$-SAT, for $6\le k\le 9$.}\label{T-add}
\end{table}

Unlike in the constructions of the previous sections,
the local distributions used for some of the additional
clause types do not satisfy the clauses with probability~$1$.
Interestingly, for each of the three clause types
$X^2Y^4$, $X^4Y^3$ and $X^8$, the local distribution used
in our construction maximizes the probability of satisfying
the clause among all local distributions consistent with
the chosen inner product matrix~$C$.
For the clause type $X^9$ we are again able to satisfy
each clause with probability~$1$.
See Appendix~\ref{A} for details.

\subsection{Constructions using three families of variables}\label{sec:PSD-3}
We obtained further improvements by introducing a third
family of variables~$Z$.
In each of the four constructions of Tables~\ref{T-three6}-\ref{T-three9} 
we introduce three additional clause types, as shown in Table~\ref{T-add2}.

In all these four constructions (Tables~\ref{T-three6}-\ref{T-three9}), each clause contains at most one variable from the family~$Z$. As such, the family $Z$ need not consist of an arbitrarily large collection of variables. Rather, we let the family~$Z$ consist of a \emph{single} variable. This modification changes neither the value of the exhibited SDP solution nor the weight of the clauses that can be simultaneously satisfied.

\begin{table}
\begin{center}
\begin{tabular}{c@{\qquad}c@{\qquad}c}
\toprule
$k$ & Additional clause types & Upper bound on $\alpha_k$ \\
\midrule
6 & $Z,\ X^4YZ,\ X^5\neg Z$   & $0.87401443$ \\
7 & $Z,\ X^5\neg Z,\ X^6Z$    & $0.87397813$ \\
8 & $Y^3Z,\ Y^3\neg Z,\ X^8$  & $0.87354080$ \\
9 & $Y^3Z,\ Y^3\neg Z,\ X^9$  & $0.87337371$ \\
\bottomrule
\end{tabular}
\end{center}
\caption{Additional clause types added to the MAX 5-SAT construction following the addition of a third family $Z$ of variables. The resulting constructions yield further improved constructions for MAX $k$-SAT, for $6\le k\le 9$.}\label{T-add2}
\end{table}

In all four constructions, the vector~$\bv_Z$ is a linear
combination of $\bv_0$ and $\bv_Y$.
More specifically, for $k=6,7,8,9$, respectively, we use
\[
\bv_Z \EQ
\begin{cases}
\frac3{20}(\bv_Y-\bv_0) \;, & k=6 \;,\\[4pt]
\frac1{150}(21\bv_Y-23\bv_0) \;, & k=7 \;,\\[4pt]
\mathbf{0} \;, & k=8,9 \;.
\end{cases}
\]
Thus, the inner product matrix~$C$ is still of rank~$2$,
as in all our two-family constructions.\footnote{We also numerically explored having the inner product matrix $C$ have rank greater than $2$, but the numerical improvements were rather slight (much less than $10^{-6}$).}

The constructions for $k=8,9$ use an additional idea.
The two clause types $Y^3Z$ and $Y^3\neg Z$ have equal
total weights.
If a clause $Y_1\vee Y_2\vee Y_3$ is satisfied, then both
clauses
\[
Y_1\vee Y_2\vee Y_3\vee Z
\quad,\quad
Y_1\vee Y_2\vee Y_3\vee\neg Z
\]
are satisfied.
Otherwise, exactly one of them is satisfied.
Consequently, the total weight of the satisfied clauses
of these two types does not depend on the values assigned
to the $Z$ variables. In our SDP solution, we choose local distributions that
satisfy all clauses of both types with probability~$1$.

\subsection{Analysis of the integral optima}

A key property used in the analysis of our constructions is 
that the computation of the maximum weight of clauses
that can be simultaneously satisfied, in the limit as
$n\to\infty$, can be reduced to univariate optimization.
This is true even for the constructions using three
families of variables.
For the three-family constructions with $k=6,7$, the
objective function is affine in the fraction of the~$Z$ variables set to~$1$, and hence it suffices to
consider the two cases in which either all or none of
the~$Z$ variables are set to~$1$.
For $k=8,9$, the objective function does not depend
on this fraction at all.
In all cases, the remaining bivariate optimization
problem can be reduced to a univariate optimization
problem, in a manner similar to the arguments in
Sections \ref{S-HZ} and \ref{S-MAX-5-SAT}.
This is explained in Appendix~\ref{A}.

\subsection{Beyond the basic SDP}

We highlight that all the inner product matrices used in our constructions
can be realized by the first and second moments of a
global distribution over Boolean assignments.
In fact, in all cases such a distribution can be obtained
as a mixture of just two product distributions. As a result, our integrality gap construction holds for a much stronger family of relaxations that the basic SDP. We describe this property in detail in Appendix~\ref{A}. We remark that the local distributions themselves cannot all
be marginals of a single global distribution, as otherwise
there would be no integrality gap.

\section{AI Disclosure}\label{sec:disclosure}

In 2021, before the age of modern AI, after completing the conference version of \cite{BHPZ25NAE}, we investigated whether the uncorrelated integrality gap instance of Halperin and Zwick \cite{HZ01} could lead to even better SDP integrality gaps by considering more complex clause types. Using SageMath~\cite{sagemath}, we wrote code which computed integrality gaps for various families of clause types, including families which included $\neg X$, $X_1 \vee X_2 \vee X_3 \vee \neg Y$, and $Y_1 \vee Y_2 \vee Y_3 \vee Y_4 \vee Y_5$. However, we did not think to include $Y_1 \vee Y_2 \vee \neg Y_3$ and gave up the search after failing to have the integrality ratio drop below $\frac78$.

In late 2026, using ChatGPT 6 Astra, we revisited this problem and found that adding $Y_1 \vee Y_2 \vee \neg Y_3$ could drop the integrality ratio below $\frac78$. We also verified this finding with a one-line modification to our old code. We also used ChatGPT 6 Astra to investigate many other families of clause types, which are documented in Section~\ref{S-MAX-6-SAT}. We used a further hand-modification of our old code to check that many of the numerical claims in Section~\ref{S-MAX-6-SAT} and Appendix~\ref{A} are accurate.

With further prompting, ChatGPT 6 Astra also found that ``correlated'' integrality gaps for MAX-SAT can also be analyzed, leading to an integrality ratio less than $0.8749$ for MAX-$4$-SAT. As the proof is %
considerably more complex, we plan to document it carefully in a future version of this manuscript.

Most of the writing in this manuscript was done by the authors, and we take full responsibility for all the content of this work. ChatGPT provided preliminary versions of Section~\ref{S-MAX-6-SAT} and of Appendix~\ref{A}, which underwent substantial edits, and assisted by giving feedback on the exposition as well as in creation of the tables and of Figure~\ref{F-max5-ratio}.

\section{Concluding remarks}

We proved that MAX SAT does not have a $\frac78$-approximation algorithm when clauses of all lengths are allowed, assuming UGC and $\mathsf{P} \neq \mathsf{NP}$. This should be contrasted with the fact that MAX SAT does have a $\frac78$-approximation algorithm when all clauses are of length at most 3, or when all clauses are of length at least 3. The mixture of these cases is what makes the problem hard.

 In the next version of this paper we will present more complicated constructions that obtain a slightly improved upper bound for MAX SAT and an upper bound below $\frac78$ also for MAX 4-SAT.

\paragraph{Acknowledgments.} JB thanks Euiwoong Lee, Venkatesan Guruswami, and Mahdi Cheraghchi for helpful discussions.

\newpage
\clearpage
\appendix
\pagestyle{appendix}

\section{Constructions for MAX $k$-SAT, $k=6,7,8,9$}\label{A}

In this appendix, we give complete descriptions of the
constructions mentioned in Section~\ref{S-MAX-6-SAT}
and explain how the upper bounds are obtained. Appendix~\ref{app:additional} gives further observations about our integrality gap constructions.

\subsection{General properties of the constructions}

All the constructions in this appendix are based on the
same general framework used in Sections~\ref{S-HZ}, \ref{S-MAX-5-SAT} and~\ref{S-MAX-6-SAT}. We use $r\le 3$ families of variables
$P_1,\ldots,P_r$, each containing $n$ variables, and $m$
different clause types. A clause of type~$t$ contains
$p_{ti}$ positive literals and $q_{ti}$ negative literals
from family~$P_i$, for $1\le i\le r$.
All the variables appearing in a clause are distinct.
All clauses of type~$t$ have the same weight and their
total weight is~$w_t$.
Let~$I_n$ be the resulting instance, for sufficiently
large~$n$.

The following two lemmas generalize properties that we used in Sections~\ref{S-HZ}, \ref{S-MAX-5-SAT} and~\ref{S-MAX-6-SAT}.

\begin{lemma}\label{L-family-SDP}
Let $C=(c_{ij})_{0\le i,j\le r}$ be a positive
semidefinite matrix with $c_{00}=1$ and $c_{ii}\le1$
for $1\le i\le r$.
Suppose that for each clause type~$t$ we have a local
distribution that satisfies a clause of this type with
probability~$s_t$ and is consistent with~$C$, i.e.,
\[
\Pr[X=1] \EQ \frac{1-c_{0i}}2
\quad,\quad
\Pr[X\ne Y] \EQ \frac{1-c_{ij}}2
\;,
\]
whenever $X$ and $Y$ are distinct variables of the clause
belonging to families $P_i$ and $P_j$, respectively.
Then
\[
\SDP(I_n) \GE \sum_{t=1}^m w_t s_t \;.
\]
\end{lemma}

\begin{proof}
As $C\succeq0$, there exist vectors
$\bv_0,\bv_{P_1},\ldots,\bv_{P_r}$
whose inner products are specified by~$C$.
For each variable~$x$ belonging to family~$P_i$,
we define
\[
\bv_x \EQ \bv_{P_i}+\sqrt{1-c_{ii}}\,\bu_x \;,
\]
where the vectors $\bu_x$ are unit vectors orthogonal
to each other and to $\bv_0,\bv_{P_1},\ldots,\bv_{P_r}$.

It is easy to check that $\bv_0$ and all the vectors
$\bv_x$ are unit vectors and that
\[
\bv_0\cdot\bv_x \EQ c_{0i}
\quad,\quad
\bv_x\cdot\bv_y \EQ c_{ij} \;,
\]
whenever $x$ and $y$ are distinct variables belonging
to families $P_i$ and $P_j$, respectively.
Thus, these vectors, together with the given local
distributions, constitute a feasible solution of the
SDP relaxation of value $\sum_{t=1}^m w_t s_t$.
\end{proof}

\begin{lemma}\label{L-family-OPT}
For the instance~$I_n$ described above,
\[
\lim_{n\to\infty}\OPT(I_n)
\EQ
\max_{x_1,\ldots,x_r\in[0,1]} F(x_1,\ldots,x_r) \;,
\]
where
\[
F(x_1,\ldots,x_r)
\EQ
\sum_{t=1}^m w_t
\left(
1-\prod_{i=1}^r(1-x_i)^{p_{ti}}x_i^{q_{ti}}
\right) \;.
\]
\end{lemma}

\begin{proof}
Consider a Boolean assignment in which exactly $n_i$
variables of family~$P_i$ are set to~$1$, and let
$x_i=n_i/n$, for $1\le i\le r$.

The total number of clauses of type~$t$ is
\[
\prod_{i=1}^r
\binom{n}{p_{ti}}\binom{n-p_{ti}}{q_{ti}}\;,
\]
while the number of clauses of type~$t$ not satisfied
by the assignment is
\[
\prod_{i=1}^r
\binom{n-n_i}{p_{ti}}\binom{n_i}{q_{ti}}\;.
\]
Thus, the total weight of the clauses satisfied by
the assignment is
\[
F_n(x_1,\ldots,x_r) \EQ
\sum_{t=1}^m w_t
\left(
1-\prod_{i=1}^r
\frac{(n-n_i)_{p_{ti}}(n_i)_{q_{ti}}}
     {(n)_{p_{ti}+q_{ti}}}
\right)\;,
\]
where $(n)_k= n!/(n-k)!$. For any fixed nonnegative integers $p,q$, we have
\[
\frac{(n-n_i)_p(n_i)_q}{(n)_{p+q}}
\EQ
\frac{
\ds\prod_{j=0}^{p-1}\left(1-x_i-\frac jn\right)
\prod_{j=0}^{q-1}\left(x_i-\frac jn\right)
}{
\ds\prod_{j=0}^{p+q-1}\left(1-\frac jn\right)
}
\EQ
(1-x_i)^px_i^q+O\left(\frac1n\right)\;,
\]
uniformly over $x_i\in[0,1]$.
Consequently,
\[
F_n(x_1,\ldots,x_r)
\EQ F(x_1,\ldots,x_r)+O\left(\frac1n\right)\;,
\]
uniformly over all Boolean assignments. Since
\[
\OPT(I_n) \EQ
\max_{x_1,\ldots,x_r\in\{0,\frac1n,\ldots,1\}}
F_n(x_1,\ldots,x_r)\;,
\]
the result follows from the continuity of~$F$ and
the fact that the grid
$\{0,\frac1n,\ldots,1\}^r$ is dense in $[0,1]^r$
as $n\to\infty$.
\end{proof}
Combining Lemmas~\ref{L-family-SDP} and~\ref{L-family-OPT},
we get the following corollary.

\begin{corollary} If the length of all clauses used in a construction belong to~$K$ then
    \[
\alpha_K \LE
\frac{\ds
\max_{x_1,\ldots,x_r\in[0,1]}
F(x_1,\ldots,x_r)}
{\ds\sum_{t=1}^m w_t s_t}
\;,
\]
provided that the denominator is positive.
\end{corollary}

\subsection{Checking SDP feasability}

Our goal is to now analyze the feasibility of the basic SDP for the constructions presented in Tables~\ref{T-two6}--\ref{T-three9} of Appendix~\ref{A}. First, note that the inner product matrices presented in these tables are PSD. For Tables~\ref{T-two6}--\ref{T-two9}, the inner product matrix is identical to the one appear in Table~\ref{T-five}, so there is nothing to check. For Tables~\ref{T-three6}--\ref{T-three9}, the fact they are PSD follows from the choices of $\bv_Z$ presented in Section~\ref{sec:PSD-3}.

The next step is to verify that there are local distributions on the clauses consistent with these inner product matrices. All the constructions in Tables~\ref{T-two6}--\ref{T-three9} for the first four clause types use identical local distributions as that of Table~\ref{T-five}(c). Thus, it suffices to construct suitable local distributions for the other clause types. These local distributions are described in full in their respective tables. However, to keep the distributions more succinct, we only describe the distributions up to permutation of variables of the same type. The number of permutations needed to be considered simultaneously is indicated by the column indexed by $m$.

\subsection{Computing the optimal integral assignment}

Our goal is now to verify the optimal assignments to the constructions given in Tables~\ref{T-two6}--\ref{T-three9} correspond to the ratios given in the respective tables. 

Let $x,y,z$ be the fractions of variables from families
$X,Y,Z$, respectively, that are set to~$1$. Let 
\[\bar{x}=1-x \;,\]
be the fraction of the variables from~$X$ that are set to~$0$. 
Recall that all our constructions retain the four clause types of the MAX 5-SAT construction given in Table~\ref{T-five}. We thus let
\[
\begin{aligned}
F_0(x,y) \EQ {}&
w_1\bar{x}+w_2\bigl(1-y(1-y)^2\bigr)
+w_3\bigl(1-\bar{x}^3y\bigr)
+w_4\bigl(1-(1-y)^5\bigr)\;,
\end{aligned}
\]
for the contribution of these four clause types
to the limiting weight of the satisfied clauses.
Similarly, the contribution of these four clause
types to the value of the SDP solution is
\[
D_0 \EQ w_1\frac{1+a}{2}+w_2+w_3+w_4\;.
\]
The weights $w_1,\ldots,w_4$ are the weights
of these clause types in the construction considered.

\subsubsection{Verifying Tables \ref{T-two6}--\ref{T-two9}}

In each of the four constructions using two families,
we introduce a clause type of the form~$X^pY^q$.
We thus have
\[
F(x,y) \EQ F_0(x,y)+w_5\bigl(1-\bar{x}^p(1-y)^q\bigr)
\]
and
\[
D \EQ D_0+w_5s_5\;,
\]
where $s_5$ is the probability that the local distribution
satisfies a clause of the additional type.

For every fixed~$y$, the function~$F$ is a concave
function of~$\bar{x}$.
Its derivative with respect to~$\bar{x}$ is
\[
w_1-3w_3y\bar{x}^2-pw_5(1-y)^q\bar{x}^{p-1}\;.
\]
If
\[
w_1\GE 3w_3y+pw_5(1-y)^q\;,
\]
the optimal choice is $\bar{x}=1$, or equivalently $x=0$.
Otherwise, the optimal choice is the unique solution
in $(0,1)$ of the equation
\[
3w_3y\bar{x}^2+pw_5(1-y)^q\bar{x}^{p-1}\EQ w_1\;.
\]
We are thus left with a univariate optimization
problem over $y\in[0,1]$.

\subsubsection{Verifying Tables \ref{T-three6}--\ref{T-three9}}

For the three-family constructions for $k=6,7,8,9$,
the function $F$, besides being made trivariate, is modified as follows.

For $k=6$ we have that
\[
\begin{aligned}
F(x,y,z) \EQ {}& F_0(x,y)+w_5z +w_6\bigl(1-\bar{x}^4(1-y)(1-z)\bigr)
+w_7(1-\bar{x}^5z)\;,
\end{aligned}
\]
and
\[
D \EQ D_0+\frac{11}{20}w_5+w_6+w_7\;.
\]

For $k=7$ we have that
\[
\begin{aligned}
F(x,y,z) \EQ {}& F_0(x,y)+w_5z
+w_6(1-\bar{x}^5z)+w_7\bigl(1-\bar{x}^6(1-z)\bigr)\;,
\end{aligned}
\]
and
\[
D \EQ D_0+\frac{83}{150}w_5+w_6+w_7\;.
\]

In both cases, for fixed values of $x,y$, the function
$F(x,y,z)$ is an affine function of~$z$.
It is thus maximized either at $z=0$ or at $z=1$.
In particular,
\[
\max_{x,y,z\in[0,1]}F(x,y,z)
\EQ
\max\left\{
\max_{x,y\in[0,1]}F(x,y,0),
\max_{x,y\in[0,1]}F(x,y,1)
\right\}\;.
\]
Furthermore, for $z=0$ and $z=1$, the objective function
has the same form as in the two-family constructions.
Thus, by optimizing over~$x$ for each fixed~$y$,
we reduce the computation of the maximum to two
univariate optimization problems.

For $k=8,9$, using the fact that $w_5=w_6$, we have that
\[
F(x,y,z) \EQ F_0(x,y)
+w_5\bigl(2-(1-y)^3\bigr)
+w_7(1-\bar{x}^k)\;,
\]
and
\[
D \EQ D_0+2w_5+w_7s_7\;,
\]
where $s_7=\frac{23}{18}-\frac a2$ for $k=8$ and
$s_7=1$ for $k=9$.
The function~$F$ does not depend on~$z$.
As before, we can optimize over~$x$ for every fixed~$y$.
The optimal choice is $\bar{x}=1$ if
$w_1\ge3w_3y+kw_7$, and otherwise it is given by the
unique solution in $(0,1)$ of the equation
\[
3w_3y\bar{x}^2+kw_7\bar{x}^{k-1}\EQ w_1\;.
\]
Thus, again, we are left with just one univariate
optimization problem.

\subsection{Additional observations}\label{app:additional}

Now that we have completed the analysis of Tables~\ref{T-two6}--\ref{T-three9} as well as proved Theorem~\ref{thm:main}(b), we detail a few additional observations concerning our integrality gap constructions. In particular, Appendix~\ref{subsec:global} shows that our integrality gap holds in much stronger contexts than the basic SDP. Appendix~\ref{subsec:SDP-local} demonstrates that the distribution of local assignments given in Tables~\ref{T-two6}--\ref{T-three9} are optimal.

\subsubsection{Global realizability of the inner products}\label{subsec:global}
An interesting property of all our constructions is that
there exists a global distribution over Boolean assignments
whose first and second moments are consistent with all the
inner products of our SDP solution. This is much stronger than merely being a feasible solution to the basic SDP.
Furthermore, in each case such a distribution can be
obtained as a mixture of just two product distributions.

To see this, recall that in all our constructions we have
$\bv_0=(1,0)$.
Let $\bv_{P_i}=(a_i,b_i)$, for $1\le i\le r$.
Let~$T$ be a random variable satisfying
$\Exp[T]=0$ and $\Exp[T^2]=1$.
Conditioned on~$T$, set each variable of family~$P_i$
to~$1$, independently, with probability
\[
p_i(T) \EQ \frac{1-a_i-b_iT}{2}\;,
\]
assuming that $p_i(T)\in[0,1]$ for every~$i$.

For every variable~$X$ belonging to family~$P_i$, we have
\[
\Pr[X=1]
\EQ \Exp[p_i(T)]
\EQ \frac{1-a_i}{2}
\EQ \frac{1-c_{0i}}2\;.
\]
Similarly, for every two distinct variables $X,Y$
belonging to families $P_i,P_j$, respectively, we have
\[
\begin{aligned}
\Pr[X\ne Y]
&\EQ \Exp[p_i(T)(1-p_j(T))+(1-p_i(T))p_j(T)]\\
&\EQ \frac12\cdot\Exp[1-(a_i+b_iT)(a_j+b_jT)]\\
&\EQ \frac{1-a_ia_j-b_ib_j}{2}\\
&\EQ \frac{1-c_{ij}}2\;.
\end{aligned}
\]
Thus, this global distribution is consistent with all
the inner products.

For the construction of Section~\ref{S-HZ} we can choose
$T=-1$ and $T=1$, each with probability~$1/2$.
The resulting probabilities of setting variables of the
two families to~$1$ are given below.
Here $a=(4+\sqrt{17})/15$.

\begin{center}
\begin{tabular}{c@{\qquad}cc}
\toprule
Probability & $X$ & $Y$ \\
\midrule
\noalign{\vskip -5pt}
$\frac12$ & $\frac16+\frac a2$ & $\frac35$ \rule{0pt}{16pt}\\[5pt]
$\frac12$ & $\frac56-\frac{3a}{2}$ & $\frac15$ \\[2pt]
\bottomrule
\end{tabular}
\end{center}

For all our constructions for MAX $k$-SAT, $5\le k\le9$,
we can choose $T=-\sqrt2$ with probability~$1/3$ and
$T=1/\sqrt2$ with probability~$2/3$.
The resulting probabilities are given below, where
$a=(2+\sqrt{13})/9$.

\begin{center}
\begin{tabular}{c@{\qquad}ccccc}
\toprule
Probability & $X$ & $\quad Y\quad$ & $Z$ ($k=6$) & $Z$ ($k=7$)
                      & $Z$ ($k=8,9$) \\
\midrule
\noalign{\vskip -5pt}
$\frac13$ & $\frac16+\frac a2$ & $\frac23$
          & $\frac35$ & $\frac35$ & $\frac12$
          \rule{0pt}{16pt}\\[5pt]
$\frac23$ & $\frac23-a$ & $\frac16$
          & $\frac{21}{40}$ & $\frac{53}{100}$ & $\frac12$ \\[2pt]
\bottomrule
\end{tabular}
\end{center}

For the two-family constructions, we use only the
columns corresponding to~$X$ and~$Y$.
For the three-family constructions, we use in addition
the column corresponding to~$Z$ for the appropriate
value of~$k$.
Note that all the probabilities are indeed in~$[0,1]$.

Although the marginals of such a global distribution
on the variables of each clause are consistent with
the inner products of our SDP solution, using these
marginals as the local distributions would decrease
the value of the objective function. Indeed, the
resulting objective value would be the expected total
weight of the clauses satisfied by a Boolean assignment
chosen according to the global distribution, and hence
could not exceed $\OPT(I_n)$.

Our constructions thus establish the same upper bounds
on the integrality ratios even if we strengthen the
basic SDP relaxation by requiring that all the inner
products be consistent with a single global distribution
over Boolean assignments.

\subsubsection{Optimality of the SDP local distributions}\label{subsec:SDP-local}

In Tables~\ref{T-two6}--\ref{T-three9}, all our local distributions satisfy the non-unit clauses with probability~$1$, except for the three clause types listed below.

\begin{center}
\begin{tabular}{c@{\qquad}c@{\qquad}c@{\qquad}c}
\toprule
Clause type & Satisfaction probability & Problem & Constructions \\
\midrule
$X^2Y^4$ & $\frac{29}{18}-a$
         & MAX 6-SAT & Table~\ref{T-two6} \\[5pt]
$X^4Y^3$ & $\frac{20}{9}-2a$
         & MAX 7-SAT & Table~\ref{T-two7} \\[5pt]
$X^8$    & $\frac{23}{18}-\frac a2$
         & MAX 8-SAT & Tables~\ref{T-two8} and~\ref{T-three8} \\
\bottomrule
\end{tabular}
\end{center}

In particular, our three-family constructions for
MAX 6-SAT (Table~\ref{T-three6}), MAX 7-SAT (Table~\ref{T-three7}) and MAX 9-SAT (Table~\ref{T-three9}) satisfy all their
non-unit clauses with probability~$1$.

We next show that, for our choice of the inner product
matrix~$C$, it is not possible to satisfy clauses of
these three exceptional types with higher probabilities.
Recall that the inner products specified by~$C$ imply
the following probabilities:
\[
\begin{array}{lll}
\ds\Pr[X=1]=\frac{1-a}{2}\;, &&
\ds\Pr[Y=1]=\frac13\;,\\[6pt]
\ds\Pr[X_1=X_2=1]=\frac7{18}-\frac a2\;, &&
\ds\Pr[Y_1=Y_2=1]=\frac16\;,\\[6pt]
\ds\Pr[X=Y=1]=\frac19\;.
\end{array}
\]

Consider a clause of type $X^pY^q$, where $q\ge1$.
Distinguishing one of the $Y$ variables, say $Y_1$,
and using the union bound, the probability of satisfying
the clause is at most
\[
\begin{aligned}
&\Pr[Y_1=1]
+\sum_{i=1}^p\Pr[X_i=1,Y_1=0]
+\sum_{j=2}^q\Pr[Y_j=1,Y_1=0]\\
&\qquad\EQ
\frac13+
p\left(\frac{1-a}{2}-\frac19\right)
+(q-1)\left(\frac13-\frac16\right)\\
&\qquad\EQ
p\left(\frac7{18}-\frac a2\right)+\frac{q+1}{6}\;.
\end{aligned}
\]
For $(p,q)=(2,4)$ and $(p,q)=(4,3)$, this gives exactly
the probabilities attained by the corresponding local
distributions in our constructions.

Similarly, the probability of satisfying a clause
of type~$X^p$ is at most
\[
\begin{aligned}
\Pr[X_1=1]+\sum_{i=2}^p\Pr[X_i=1,X_1=0]
&\EQ \frac{1-a}{2}+\frac{p-1}{9}\;.
\end{aligned}
\]
For $p=8$ this gives exactly the probability attained
by the local distribution in our construction.
For $p=9$ this upper bound exceeds~$1$, and indeed we
are able to satisfy the clause with probability~$1$.

\begin{table}[p]
\centering
\setlength{\tabcolsep}{3pt}
\setlength{\arraycolsep}{3pt}
\[\begin{array}{c}
\begin{array}{c@{\quad}c@{\quad}c}
\toprule
\text{Type}&\text{Clause}&\text{Total weight}\\
\midrule
1&\neg X& w_1=238254563211 \\
2&Y_1\vee Y_2\vee\neg Y_3& w_2=145954228045 \\
3&X_1\vee X_2\vee X_3\vee\neg Y& w_3=423099816689 \\
4&Y_1\vee Y_2\vee Y_3\vee Y_4\vee Y_5& w_4=187593266604 \\
5&X_1\vee X_2\vee Y_1\vee Y_2\vee Y_3\vee Y_4& w_5=5098125452 \\
\bottomrule
\end{array}\\ \\
\text{(a) Variable families and clause types.}
\end{array}\]
\[\begin{array}{c}
C \EQ \begin{pmatrix}
1&a&\frac13\\
a&\frac59&a-\frac29\\
\frac13&a-\frac29&\frac13
\end{pmatrix}\\[25pt]
a \EQ \frac{2+\sqrt{13}}9\approx0.622839\\[10pt]
\text{(b) Inner products.}
\end{array}\]
\vspace{8pt}
\begin{tabular}[t]{@{}cccccccc@{}}
\multicolumn{8}{c}{$\boldsymbol{X_1\vee X_2\vee Y_1\vee Y_2\vee Y_3\vee Y_4}$}\\[2mm]
\toprule
$X_1$ & $X_2$ & $Y_1$ & $Y_2$ & $Y_3$ & $Y_4$ & $m$ & Probability \\
\midrule
0 & 0 & 0 & 0 & 0 & 0 & 1 & $a-\frac{11}{18}$ \\[4pt]
0 & 0 & 1 & 0 & 0 & 0 & 4 & $\frac16$ \\[4pt]
0 & 0 & 1 & 1 & 1 & 1 & 1 & $\frac13-\frac a2$ \\[4pt]
1 & 0 & 0 & 0 & 0 & 0 & 2 & $\frac7{18}-\frac a2$ \\[4pt]
1 & 0 & 1 & 1 & 1 & 1 & 2 & $\frac a2-\frac5{18}$ \\[4pt]
1 & 1 & 1 & 1 & 1 & 1 & 1 & $\frac7{18}-\frac a2$ \\
\bottomrule
\end{tabular}\par\vspace{6pt}{\footnotesize Types 1--4 use Table~\ref{T-five}(c); type 5 is displayed above.}

\par\vspace{10pt}
(c) Local distributions.
\[
\begin{array}{c}
x_*\EQ 0\\
y_*\approx 0.155768505830\\[10pt]
\text{(d) Optimal fractions.}
\end{array}
\hspace*{1cm}
\begin{array}{c}
\alpha_{\{1,3,4,5,6\}} \LT 0.8741802578\\[10pt]
\text{(e) Upper bound.}
\end{array}
\]
\caption{Two-family construction for MAX $\{1,3,4,5,6\}$-SAT.}\label{T-two6}
\end{table}

\begin{table}[p]
\centering
\setlength{\tabcolsep}{3pt}
\setlength{\arraycolsep}{3pt}
\[\begin{array}{c}
\begin{array}{c@{\quad}c@{\quad}c}
\toprule
\text{Type}&\text{Clause}&\text{Total weight}\\
\midrule
1&\neg X& w_1=237434632265 \\
2&Y_1\vee Y_2\vee\neg Y_3& w_2=158639833035 \\
3&X_1\vee X_2\vee X_3\vee\neg Y& w_3=415184707100 \\
4&Y_1\vee Y_2\vee Y_3\vee Y_4\vee Y_5& w_4=176774431293 \\
5&X_1\vee X_2\vee X_3\vee X_4\vee Y_1\vee Y_2\vee Y_3& w_5=11966396307 \\
\bottomrule
\end{array}\\ \\
\text{(a) Variable families and clause types.}
\end{array}\]
\[\begin{array}{c}
C \EQ \begin{pmatrix}
1&a&\frac13\\
a&\frac59&a-\frac29\\
\frac13&a-\frac29&\frac13
\end{pmatrix}\\[25pt]
a \EQ \frac{2+\sqrt{13}}9\approx0.622839\\[10pt]
\text{(b) Inner products.}
\end{array}\]
\vspace{8pt}
\begin{tabular}[t]{@{}ccccccccc@{}}
\multicolumn{9}{c}{$\boldsymbol{X_1\vee X_2\vee X_3\vee X_4\vee Y_1\vee Y_2\vee Y_3}$}\\[2mm]
\toprule
$X_1$ & $X_2$ & $X_3$ & $X_4$ & $Y_1$ & $Y_2$ & $Y_3$ & $m$ & Probability \\
\midrule
0 & 0 & 0 & 0 & 0 & 0 & 0 & 1 & $2a-\frac{11}{9}$ \\[4pt]
0 & 0 & 0 & 0 & 1 & 0 & 0 & 3 & $\frac16$ \\[4pt]
1 & 0 & 0 & 0 & 0 & 0 & 0 & 4 & $\frac7{18}-\frac a2$ \\[4pt]
1 & 0 & 0 & 0 & 1 & 1 & 1 & 4 & $\frac16-\frac a4$ \\[4pt]
1 & 1 & 1 & 0 & 1 & 1 & 1 & 4 & $\frac{3a}{4}-\frac49$ \\[4pt]
1 & 1 & 1 & 1 & 1 & 1 & 1 & 1 & $\frac{23}{18}-2a$ \\
\bottomrule
\end{tabular}\par\vspace{6pt}{
Types 1--4 use Table~\ref{T-five}(c); type 5 is displayed above.}

\par\vspace{10pt}
(c) Local distributions.
\[
\begin{array}{c}
x_*\approx 0.495988502441\\
y_*\approx 0.750111157451\\[10pt]
\text{(d) Optimal fractions.}
\end{array}
\hspace*{1cm}
\begin{array}{c}
\alpha_{\{1,3,4,5,7\}} \LT 0.8741313135\\[10pt]
\text{(e) Upper bound.}
\end{array}
\]
\caption{Two-family construction for MAX $\{1,3,4,5,7\}$-SAT.}\label{T-two7}
\end{table}

\begin{table}[p]
\centering
\setlength{\tabcolsep}{3pt}
\setlength{\arraycolsep}{3pt}
\[\begin{array}{c}
\begin{array}{c@{\quad}c@{\quad}c}
\toprule
\text{Type}&\text{Clause}&\text{Total weight}\\
\midrule
1&\neg X& w_1=238575093527 \\
2&Y_1\vee Y_2\vee\neg Y_3& w_2=163783211381 \\
3&X_1\vee X_2\vee X_3\vee\neg Y& w_3=414275964043 \\
4&Y_1\vee Y_2\vee Y_3\vee Y_4\vee Y_5& w_4=174164703128 \\
5&X_1\vee\cdots\vee X_8& w_5=9201027921 \\
\bottomrule
\end{array}\\ \\
\text{(a) Variable families and clause types.}
\end{array}\]
\[\begin{array}{c}
C \EQ \begin{pmatrix}
1&a&\frac13\\
a&\frac59&a-\frac29\\
\frac13&a-\frac29&\frac13
\end{pmatrix}\\[25pt]
a \EQ \frac{2+\sqrt{13}}9\approx0.622839\\[10pt]
\text{(b) Inner products.}
\end{array}\]
\vspace{8pt}
\begin{tabular}[t]{@{}cccccccccc@{}}
\multicolumn{10}{c}{$\boldsymbol{X_1\vee\cdots\vee X_8}$}\\[2mm]
\toprule
$X_1$ & $X_2$ & $X_3$ & $X_4$ & $X_5$ & $X_6$ & $X_7$ & $X_8$ & $m$ & Probability \\
\midrule
0 & 0 & 0 & 0 & 0 & 0 & 0 & 0 & 1 & $\frac a2-\frac5{18}$ \\[4pt]
1 & 0 & 0 & 0 & 0 & 0 & 0 & 0 & 8 & $\frac19$ \\[4pt]
1 & 1 & 1 & 1 & 1 & 1 & 1 & 1 & 1 & $\frac7{18}-\frac a2$ \\
\bottomrule
\end{tabular}\par\vspace{6pt}{\footnotesize Types 1--4 use Table~\ref{T-five}(c); type 5 is displayed above.}

\par\vspace{10pt}
(c) Local distributions.
\[
\begin{array}{c}
x_*\approx 0.503959452829\\
y_*\approx 0.778374257119\\[10pt]
\text{(d) Optimal fractions.}
\end{array}
\hspace*{1cm}
\begin{array}{c}
\alpha_{\{1,3,4,5,8\}} \LT 0.87359549\\[10pt]
\text{(e) Upper bound.}
\end{array}
\]
\caption{Two-family construction for MAX $\{1,3,4,5,8\}$-SAT.}\label{T-two8}
\end{table}

\begin{table}[p]
\centering
\setlength{\tabcolsep}{3pt}
\setlength{\arraycolsep}{3pt}
\[\begin{array}{c}
\begin{array}{c@{\quad}c@{\quad}c}
\toprule
\text{Type}&\text{Clause}&\text{Total weight}\\
\midrule
1&\neg X& w_1=237555353144 \\
2&Y_1\vee Y_2\vee\neg Y_3& w_2=162602678253 \\
3&X_1\vee X_2\vee X_3\vee\neg Y& w_3=417167260356 \\
4&Y_1\vee Y_2\vee Y_3\vee Y_4\vee Y_5& w_4=173898624530 \\
5&X_1\vee\cdots\vee X_9& w_5=8776083718 \\
\bottomrule
\end{array}\\ \\
\text{(a) Variable families and clause types.}
\end{array}\]
\[\begin{array}{c}
C \EQ \begin{pmatrix}
1&a&\frac13\\
a&\frac59&a-\frac29\\
\frac13&a-\frac29&\frac13
\end{pmatrix}\\[25pt]
a \EQ \frac{2+\sqrt{13}}9\approx0.622839\\[10pt]
\text{(b) Inner products.}
\end{array}\]
\vspace{8pt}
\begin{tabular}[t]{@{}ccccccccccc@{}}
\multicolumn{11}{c}{$\boldsymbol{X_1\vee\cdots\vee X_9}$}\\[2mm]
\toprule
$X_1$ & $X_2$ & $X_3$ & $X_4$ & $X_5$ & $X_6$ & $X_7$ & $X_8$ & $X_9$ & $m$ & Probability \\
\midrule
1 & 0 & 0 & 0 & 0 & 0 & 0 & 0 & 0 & 9 & $\frac{1+9a}{72}$ \\[4pt]
1 & 1 & 0 & 0 & 0 & 0 & 0 & 0 & 0 & 36 & $\frac{7-9a}{504}$ \\[4pt]
1 & 1 & 1 & 1 & 1 & 1 & 1 & 1 & 1 & 1 & $\frac{3(7-9a)}{56}$ \\
\bottomrule
\end{tabular}\par\vspace{6pt}{ Types 1--4 use Table~\ref{T-five}(c); type 5 is displayed above.}

\par\vspace{10pt}
(c) Local distributions.
\[
\begin{array}{c}
x_*\approx 0.506848145344\\
y_*\approx 0.779590571564\\[10pt]
\text{(d) Optimal fractions.}
\end{array}
\hspace*{1cm}
\begin{array}{c}
\alpha_{\{1,3,4,5,9\}} \LT 0.8734561883\\[10pt]
\text{(e) Upper bound.}
\end{array}
\]
\caption{Two-family construction for MAX $\{1,3,4,5,9\}$-SAT.}\label{T-two9}
\end{table}

\begin{table}[p]
\centering
\setlength{\tabcolsep}{3pt}
\setlength{\arraycolsep}{3pt}
\[\begin{array}{c}
\begin{array}{c@{\quad}c@{\quad}c}
\toprule
\text{Type}&\text{Clause}&\text{Total weight}\\
\midrule
1&\neg X& w_1=239684681 \\
2&Y_1\vee Y_2\vee \neg Y_3& w_2=159615969 \\
3&X_1\vee X_2\vee X_3\vee \neg Y& w_3=403759494 \\
4&Y_1\vee Y_2\vee Y_3\vee Y_4\vee Y_5& w_4=173546007 \\
5&Z& w_5=23193 \\
6&X_1\vee X_2\vee X_3\vee X_4\vee Y\vee Z& w_6=12542790 \\
7&X_1\vee X_2\vee X_3\vee X_4\vee X_5\vee \neg Z& w_7=10827866 \\
\bottomrule
\end{array}\\ \\
\text{(a) Variable families and clause types.}
\end{array}%
\hspace{0.9cm}
\begin{array}{c}
C \EQ \begin{pmatrix}
1&a&\frac13&-\frac1{10}\\
a&\frac59&a-\frac29&-\frac1{30}\\
\frac13&a-\frac29&\frac13&0\\
-\frac1{10}&-\frac1{30}&0&\frac3{200}
\end{pmatrix}\\[25pt]
a \EQ \frac{2+\sqrt{13}}9\approx0.622839\\[10pt]
\text{(b) Inner products.}
\end{array}\]
\vspace{8pt}
\begin{tabular}[t]{@{}cc@{}}
\multicolumn{2}{c}{$\boldsymbol{Z}$}\\[2mm]
\toprule
$Z$ & Probability \\\midrule
0 & $\frac{9}{20}$\\[4pt]
1 & $\frac{11}{20}$\\[4pt]
\bottomrule
\end{tabular}
\par\vspace{8pt}
\begin{tabular}[t]{@{}cccccccc@{}}
\multicolumn{8}{c}{$\boldsymbol{X_1\vee X_2\vee X_3\vee X_4\vee X_5\vee \neg Z}$}\\[2mm]
\toprule
$X_1$ & $X_2$ & $X_3$ & $X_4$ & $X_5$ & $Z$ & $m$ & Probability \\\midrule
1 & 1 & 1 & 1 & 1 & 1 & $1$ & $\frac{47 - 75 a}{240}$\\[4pt]
1 & 1 & 1 & 1 & 1 & 0 & $1$ & $\frac{97 - 90 a}{540}$\\[4pt]
1 & 1 & 0 & 0 & 0 & 0 & $10$ & $\frac{29 - 45 a}{2160}$\\[4pt]
1 & 0 & 0 & 0 & 0 & 1 & $5$ & $\frac{15 a + 17}{240}$\\[4pt]
0 & 0 & 0 & 0 & 0 & 0 & $1$ & $\frac{135 a + 49}{360}$\\[4pt]
\bottomrule
\end{tabular}
\hspace{0.65cm}
\begin{tabular}[t]{@{}cccccccc@{}}
\multicolumn{8}{c}{$\boldsymbol{X_1\vee X_2\vee X_3\vee X_4\vee Y\vee Z}$}\\[2mm]
\toprule
$X_1$ & $X_2$ & $X_3$ & $X_4$ & $Y$ & $Z$ & $m$ & Probability \\\midrule
1 & 1 & 1 & 1 & 1 & 1 & $1$ & $\frac{25 - 24 a}{144}$\\[4pt]
1 & 1 & 1 & 1 & 1 & 0 & $1$ & $\frac{5 - 8 a}{32}$\\[4pt]
1 & 1 & 1 & 0 & 0 & 1 & $4$ & $\frac{449 - 720 a}{4320}$\\[4pt]
1 & 1 & 0 & 0 & 1 & 1 & $6$ & $\frac{1080 a - 643}{4320}$\\[4pt]
1 & 0 & 0 & 0 & 1 & 1 & $4$ & $\frac{41 - 60 a}{180}$\\[4pt]
1 & 0 & 0 & 0 & 0 & 0 & $4$ & $\frac{37}{480}$\\[4pt]
0 & 0 & 0 & 0 & 1 & 0 & $1$ & $\frac{120 a - 7}{480}$\\[4pt]
0 & 0 & 0 & 0 & 0 & 1 & $1$ & $\frac{360 a - 31}{540}$\\[4pt]
\bottomrule
\end{tabular}\par\vspace{6pt}{ Types 1--4 use Table~\ref{T-five}(c); types 5--7 are displayed above.}

\par\vspace{10pt}
(c) Local distributions.
\[
\begin{array}{c}
x_*\approx 0.48802838\\
y_*\approx 0.74961661\\
z_*\EQ 0\\[10pt]
\text{(d) Optimal fractions.}
\end{array}
\hspace*{1cm}
\begin{array}{c}
\alpha_{\{1,3,4,5,6\}} \LT 0.87401443\\[10pt]
\text{(e) Upper bound.}
\end{array}
\]
\caption{Three-family construction for MAX 6-SAT.}\label{T-three6}
\end{table}

\begin{table}[p]
\centering
\setlength{\tabcolsep}{3pt}
\setlength{\arraycolsep}{3pt}
\[\begin{array}{c}
\begin{array}{c@{\quad}c@{\quad}c}
\toprule
\text{Type}&\text{Clause}&\text{Total weight}\\
\midrule
1&\neg X& w_1=239299131 \\
2&Y_1\vee Y_2\vee \neg Y_3& w_2=159771179 \\
3&X_1\vee X_2\vee X_3\vee \neg Y& w_3=405998085 \\
4&Y_1\vee Y_2\vee Y_3\vee Y_4\vee Y_5& w_4=176452141 \\
5&Z& w_5=424993 \\
6&X_1\vee X_2\vee X_3\vee X_4\vee X_5\vee \neg Z& w_6=9239732 \\
7&X_1\vee X_2\vee X_3\vee X_4\vee X_5\vee X_6\vee Z& w_7=8814740 \\
\bottomrule
\end{array}\\ \\
\text{(a) Variable families and clause types.}
\end{array}
\hspace*{0.5cm}
\begin{array}{c}
C \EQ \begin{pmatrix}
1&a&\frac13&-\frac8{75}\\
a&\frac59&a-\frac29&-\frac{3a+7}{225}\\
\frac13&a-\frac29&\frac13&-\frac1{225}\\
-\frac8{75}&-\frac{3a+7}{225}&-\frac1{225}&\frac{59}{3750}
\end{pmatrix}\\[25pt]
a \EQ \frac{2+\sqrt{13}}9\approx0.622839\\[10pt]
\text{(b) Inner products.}
\end{array}\]
\vspace{8pt}
\begin{tabular}[t]{@{}cc@{}}
\multicolumn{2}{c}{$\boldsymbol{Z}$}\\[2mm]
\toprule
$Z$ & Probability \\\midrule
0 & $\frac{67}{150}$\\[4pt]
1 & $\frac{83}{150}$\\[4pt]
\bottomrule
\end{tabular}
\hspace{0.65cm}
\begin{tabular}[t]{@{}ccccccccc@{}}
\multicolumn{9}{c}{$\boldsymbol{X_1\vee X_2\vee X_3\vee X_4\vee X_5\vee X_6\vee Z}$}\\[2mm]
\toprule
$X_1$ & $X_2$ & $X_3$ & $X_4$ & $X_5$ & $X_6$ & $Z$ & $m$ & Probability \\\midrule
1 & 1 & 1 & 1 & 1 & 1 & 1 & $1$ & $\frac{371 - 561 a}{900}$\\[4pt]
1 & 1 & 1 & 1 & 0 & 0 & 1 & $15$ & $\frac{111 a - 43}{3000}$\\[4pt]
1 & 1 & 0 & 0 & 0 & 0 & 0 & $15$ & $\frac{47 - 74 a}{750}$\\[4pt]
1 & 0 & 0 & 0 & 0 & 0 & 0 & $6$ & $\frac{111 a - 37}{450}$\\[4pt]
0 & 0 & 0 & 0 & 0 & 0 & 1 & $1$ & $\frac{123 a + 641}{1800}$\\[4pt]
\bottomrule
\end{tabular}
\par\vspace{8pt}
\begin{tabular}[t]{@{}cccccccc@{}}
\multicolumn{8}{c}{$\boldsymbol{X_1\vee X_2\vee X_3\vee X_4\vee X_5\vee \neg Z}$}\\[2mm]
\toprule
$X_1$ & $X_2$ & $X_3$ & $X_4$ & $X_5$ & $Z$ & $m$ & Probability \\\midrule
1 & 1 & 1 & 1 & 1 & 1 & $1$ & $\frac{178 - 285 a}{900}$\\[4pt]
1 & 1 & 1 & 1 & 1 & 0 & $1$ & $\frac{80 - 73 a}{450}$\\[4pt]
1 & 1 & 0 & 0 & 0 & 0 & $10$ & $\frac{12 - 19 a}{900}$\\[4pt]
1 & 0 & 0 & 0 & 0 & 1 & $5$ & $\frac{57 a + 64}{900}$\\[4pt]
0 & 0 & 0 & 0 & 0 & 0 & $1$ & $\frac{168 a + 61}{450}$\\[4pt]
\bottomrule
\end{tabular}\par\vspace{6pt}{ Types 1--4 use Table~\ref{T-five}(c); types 5--7 are displayed above.}

\par\vspace{10pt}
(c) Local distributions.
\[
\begin{array}{c}
x_*\approx 0.35801543\\
y_*\approx 0.46106904\\
z_*\EQ 1\\[10pt]
\text{(d) Optimal fractions.}
\end{array}
\hspace*{1cm}
\begin{array}{c}
\alpha_{\{1,3,4,5,6,7\}} \LT 0.87397813\\[10pt]
\text{(e) Upper bound.}
\end{array}
\]
\caption{Three-family construction for MAX 7-SAT.}\label{T-three7}
\end{table}

\begin{table}[p]
\centering
\setlength{\tabcolsep}{3pt}
\setlength{\arraycolsep}{3pt}
\[\begin{array}{c}
\begin{array}{c@{\quad}c@{\quad}c}
\toprule
\text{Type}&\text{Clause}&\text{Total weight}\\
\midrule
1&\neg X& w_1=241553808 \\
2&Y_1\vee Y_2\vee \neg Y_3& w_2=149628893 \\
3&X_1\vee X_2\vee X_3\vee \neg Y& w_3=409665071 \\
4&Y_1\vee Y_2\vee Y_3\vee Y_4\vee Y_5& w_4=146388459 \\
5&Y_1\vee Y_2\vee Y_3\vee Z& w_5=20044886 \\
6&Y_1\vee Y_2\vee Y_3\vee \neg Z& w_6=20044886 \\
7&X_1\vee\cdots\vee X_8& w_7=12673996 \\
\bottomrule
\end{array}\\ \\
\text{(a) Variable families and clause types.}
\end{array}
\hspace*{0.9cm}
\begin{array}{c}
C \EQ \begin{pmatrix}
1&a&\frac13&0\\
a&\frac59&a-\frac29&0\\
\frac13&a-\frac29&\frac13&0\\
0&0&0&0
\end{pmatrix}\\[25pt]
a \EQ \frac{2+\sqrt{13}}9\approx0.622839\\[10pt]
\text{(b) Inner products.}
\end{array}\]
\vspace{8pt}
\begin{tabular}[t]{@{}ccccc@{}}
\multicolumn{5}{c}{$\boldsymbol{Y_1\vee Y_2\vee Y_3\vee Z}$}\\[2mm]
\toprule
$Y_1$ & $Y_2$ & $Y_3$ & $Z$ & Probability \\\midrule
1 & 1 & 1 & 1 & $\frac{1}{6}$\\[4pt]
1 & 0 & 0 & 0 & $\frac{1}{6}$\\[4pt]
0 & 1 & 0 & 0 & $\frac{1}{6}$\\[4pt]
0 & 0 & 1 & 0 & $\frac{1}{6}$\\[4pt]
0 & 0 & 0 & 1 & $\frac{1}{3}$\\[4pt]
\bottomrule
\end{tabular}
\hspace{0.65cm}
\begin{tabular}[t]{@{}ccccc@{}}
\multicolumn{5}{c}{$\boldsymbol{Y_1\vee Y_2\vee Y_3\vee \neg Z}$}\\[2mm]
\toprule
$Y_1$ & $Y_2$ & $Y_3$ & $Z$ & Probability \\\midrule
1 & 1 & 1 & 0 & $\frac{1}{6}$\\[4pt]
1 & 0 & 0 & 1 & $\frac{1}{6}$\\[4pt]
0 & 1 & 0 & 1 & $\frac{1}{6}$\\[4pt]
0 & 0 & 1 & 1 & $\frac{1}{6}$\\[4pt]
0 & 0 & 0 & 0 & $\frac{1}{3}$\\[4pt]
\bottomrule
\end{tabular}\par\vspace{6pt}{ Types 1--4 use Table~\ref{T-five}(c); type 7 uses Table~\ref{T-two8}(c).}

\par\vspace{10pt}
(c) Local distributions.
\[
\begin{array}{c}
x_*\approx 0.30492433\\
y_*\approx 0.39343254\\
z_* \in[0,1]\\[10pt]
\text{(d) Optimal fractions.}
\end{array}
\hspace*{1cm}
\begin{array}{c}
\alpha_{\{1,3,4,5,8\}} \LT 0.87354080\\[10pt]
\text{(e) Upper bound.}
\end{array}
\]
\caption{Three-family construction for MAX 8-SAT.}\label{T-three8}
\end{table}

\begin{table}[p]
\centering
\setlength{\tabcolsep}{3pt}
\setlength{\arraycolsep}{3pt}
\[\begin{array}{c}
\begin{array}{c@{\quad}c@{\quad}c}
\toprule
\text{Type}&\text{Clause}&\text{Total weight}\\
\midrule
1&\neg X& w_1=240470337 \\
2&Y_1\vee Y_2\vee \neg Y_3& w_2=146051502 \\
3&X_1\vee X_2\vee X_3\vee \neg Y& w_3=413060196 \\
4&Y_1\vee Y_2\vee Y_3\vee Y_4\vee Y_5& w_4=145686749 \\
5&Y_1\vee Y_2\vee Y_3\vee Z& w_5=21439142 \\
6&Y_1\vee Y_2\vee Y_3\vee \neg Z& w_6=21439142 \\
7&X_1\vee\cdots\vee X_9& w_7=11852932 \\
\bottomrule
\end{array}\\ \\
\text{(a) Variable families and clause types.}
\end{array}
\hspace*{0.9cm}
\begin{array}{c}
C \EQ \begin{pmatrix}
1&a&\frac13&0\\
a&\frac59&a-\frac29&0\\
\frac13&a-\frac29&\frac13&0\\
0&0&0&0
\end{pmatrix}\\[25pt]
a \EQ \frac{2+\sqrt{13}}9\approx0.622839\\[10pt]
\text{(b) Inner products.}
\end{array}\]
\vspace{8pt}
\begin{tabular}[t]{@{}ccccc@{}}
\multicolumn{5}{c}{$\boldsymbol{Y_1\vee Y_2\vee Y_3\vee Z}$}\\[2mm]
\toprule
$Y_1$ & $Y_2$ & $Y_3$ & $Z$ & Probability \\\midrule
1 & 1 & 1 & 1 & $\frac{1}{6}$\\[4pt]
1 & 0 & 0 & 0 & $\frac{1}{6}$\\[4pt]
0 & 1 & 0 & 0 & $\frac{1}{6}$\\[4pt]
0 & 0 & 1 & 0 & $\frac{1}{6}$\\[4pt]
0 & 0 & 0 & 1 & $\frac{1}{3}$\\[4pt]
\bottomrule
\end{tabular}
\hspace{0.65cm}
\begin{tabular}[t]{@{}ccccc@{}}
\multicolumn{5}{c}{$\boldsymbol{Y_1\vee Y_2\vee Y_3\vee \neg Z}$}\\[2mm]
\toprule
$Y_1$ & $Y_2$ & $Y_3$ & $Z$ & Probability \\\midrule
1 & 1 & 1 & 0 & $\frac{1}{6}$\\[4pt]
1 & 0 & 0 & 1 & $\frac{1}{6}$\\[4pt]
0 & 1 & 0 & 1 & $\frac{1}{6}$\\[4pt]
0 & 0 & 1 & 1 & $\frac{1}{6}$\\[4pt]
0 & 0 & 0 & 0 & $\frac{1}{3}$\\[4pt]
\bottomrule
\end{tabular}\par\vspace{6pt}{ Types 1--4 use Table~\ref{T-five}(c); type 7 uses Table~\ref{T-two9}(c).}

\par\vspace{10pt}
(c) Local distributions.
\[
\begin{array}{c}
x_*\approx 0.32644425\\
y_*\approx 0.41970144\\
z_* \in[0,1]\\[10pt]
\text{(d) Optimal fractions.}
\end{array}
\hspace*{1cm}
\begin{array}{c}
\alpha_{\{1,3,4,5,9\}} \LT 0.87337371\\[10pt]
\text{(e) Upper bound.}
\end{array}
\]
\caption{Three-family construction for MAX 9-SAT.}\label{T-three9}
\end{table}

\newpage
\clearpage
\pagestyle{plain}

\bibliography{maxsat_merged_deduplicated}
\bibliographystyle{alphaurl}

\end{document}